\documentclass[lettersize,journal]{IEEEtran}
\usepackage{amsmath,amsfonts}
\usepackage{algorithm}
\usepackage{algpseudocode}
\usepackage{array}
\usepackage[caption=false,font=normalsize,labelfont=sf,textfont=sf]{subfig}
\usepackage{textcomp}
\usepackage{stfloats}
\usepackage{url}
\usepackage{verbatim}
\usepackage{graphicx}
\usepackage{cite}
\usepackage{tabularx,booktabs}
\usepackage{diagbox}
\usepackage{multirow}
\usepackage{multicol}
\newcolumntype{C}{>{\centering\arraybackslash}X}
\usepackage{hyperref}
\usepackage{threeparttable} 

\usepackage{xcolor}
\usepackage{soul}
\usepackage{caption}
\usepackage{comment}
 
\usepackage{color}

\begin{document}

\title{MERGE - A Bimodal Audio-Lyrics Dataset For Static Music Emotion Recognition}

\author{Pedro Lima Louro, Hugo Redinho, Tiago Filipe Rodrigues Ribeiro, Ricardo Santos, Ricardo Malheiro, Renato Panda and Rui Pedro Paiva
\thanks{P. L. Louro, H. Redinho, T. F. R. Ribeiro, R. Santos, and R. P. Paiva are with the University of Coimbra, Centre for Informatics and Systems of the University of Coimbra (CISUC), Department of Informatics Engineering, and LASI. E-mail: {pedrolouro, redinho, tiago.f.ribeiro, ricardocorreia, ruipedro}@dei.uc.pt.}
\thanks{R. Malheiro is with CISUC, LASI, and Polytechnic Institute of Leiria - School of Technology and Management. E-mail: rsmal@dei.uc.pt.}
\thanks{R. Panda is with CISUC, LASI, and Ci2 - Smart Cities Research Center, Polytechnic Institute of Tomar. Email: panda@dei.uc.pt.}
}

\markboth{Journal of \LaTeX\ Class Files,~Vol.~14, No.~8, August~2021}%
{Shell \MakeLowercase{\textit{et al.}}: A Sample Article Using IEEEtran.cls for IEEE Journals}


\maketitle

\begin{abstract}
    The Music Emotion Recognition (MER) field has seen steady developments in recent years, with contributions from feature engineering, machine learning, and deep learning. The landscape has also shifted from audio-centric systems to bimodal ensembles that combine audio and lyrics. However, a lack of public, sizable and quality-controlled bimodal databases has hampered the development and improvement of bimodal audio-lyrics systems. This article proposes three new audio, lyrics, and bimodal MER research datasets, collectively referred to as MERGE, which were created using a semi-automatic approach. To comprehensively assess the proposed datasets and establish a baseline for benchmarking, we conducted several experiments for each modality, using feature engineering, machine learning, and deep learning methodologies. Additionally, we propose and validate fixed train-validation-test splits. The obtained results confirm the viability of the proposed datasets, achieving the best overall result of 81.74\% F1-score for bimodal classification.
\end{abstract}


\section{Introduction}
Music Emotion Recognition (MER) has attracted increasing awareness within the Music Information Retrieval (MIR) community. In fact, “music’s preeminent functions are social and psychological”, and so “the most useful retrieval indexes are those that facilitate searching in conformity with such social and psychological functions. Typically, such indexes will focus on stylistic, mood, and similarity information” \cite{huron10perceptual}. Several studies on music information behavior highlighted emotional content as an essential criterion for music retrieval and organization \cite{yang2011emotion}. This growing interest was also driven by the popularity of music streaming platforms and the need to organize and recommend music to their users. 

Emotion in music can be understood in three ways: i) expressed, which refers to the emotion that the composer or performer intends to convey with the musical piece; ii) induced, which relates to the emotion that is evoked in the listener as a reaction to the song; iii) and perceived, pertaining to the emotion that the listener identifies while listening to a song (which may differ from the composer's original intention and what the listener personally feels in response) \cite{gabrielsson2001emotion}.

Over the years, several methodologies have been proposed, resulting in a substantial body of research on various aspects of MER, including emotion perception \cite{zhang2022attention}, induction \cite{paukner2025induced}, and expression \cite{Oh2024AudioFeatures}. Among these, the focus of this article is on \textit{emotion perception}. 

In terms of emotion perception, a significant portion of the studies address static MER (i.e., the identification of the predominant emotion perceived in a song), either regarding emotions as discrete labels \cite{panda2020NovelAudioFeatures, agrawal2021TransformerbasedApproachMusic}) or continuous arousal-valence (AV) values \cite{zhang2022attention}. Besides static MER, other problems have been studied, e.g., Music Emotion Variation Detection (MEVD), where emotion fluctuations throughout songs are analyzed \cite{orjesek2022EndtoendMusicEmotion}, multi-label emotion classification, where each song is assigned several emotion tags \cite{huAdvancingMusicEmotion2025}, and emotion-based automatic playlist generation \cite{donnelly2022mood}, among others. In this work, our focus is on \textit{static MER}, both using discrete emotion labels (namely Russell's emotion quadrants) and continuous AV values \cite{russell1980CircumplexModelAffect}).

MER problems have been primarily tackled using audio as the only information source. Fewer works have employed song lyrics, either alone \cite{malheiro2018EmotionallyRelevantFeaturesClassification} or following bimodal audio-lyrics strategies \cite{huAdvancingMusicEmotion2025}. It is well known that employing audio and lyrics enables the exploitation of synergies that result from the combination of the information conveyed by each source. For example, information regarding music arousal (i.e., the energy level of a musical piece) is captured mainly from the acoustic component of the music. On the other hand, the lyrical counterpart is instrumental in valence perception (i.e., how positive or negative the emotions conveyed in a song are). Although most MER systems are based on audio-only data, several psychological studies confirm the importance of lyrics in conveying semantic information. According to Juslin and Laukka \cite{Juslin01092004}, 29\% of people mention that lyrics are essential to how music conveys emotions. Additionally, Besson et al. \cite{besson1998independence} have demonstrated that part of the semantic information of songs resides exclusively in the lyrics. Audio-lyrics bimodal systems, herein referred to simply as bimodal systems, emerge naturally, and several studies have confirmed that the combination of these information sources improves MER classification results \cite{hu2010ImprovingMusicMood}. 

To support the research on static MER, several datasets have been proposed over the years. Most current datasets employ audio data, and bimodal datasets are scarce. Moreover, quality and sizable datasets are crucial to fully fulfilling the promise entailed in bimodal MER approaches. Since emotion perception is inherently subjective, creating quality and sizable datasets is a challenging and labor-intensive task, prone to disagreement among annotators \cite{yang2011emotion}. Hence, several limitations have been identified in static MER datasets (either audio-only or bimodal): limited size, lack of diversity in the music genres and styles covered, noisy annotations, noisy samples, lack of public availability, and the use of emotion taxonomies without scientific support \cite{laurier2011AutomaticClassificationMusical, panda2020NovelAudioFeatures}. These shortcomings, particularly regarding dataset scale, quality, and information source (typically audio-only), hinder the advancement of MER research.

To partly overcome these limitations, we propose a new bimodal static MER dataset, called MERGE\footnote{MERGE is the acronym of “Music Emotion Recognition next GEneration”, a research project funded by the
Portuguese Science Foundation.}, containing 2216 bimodal samples, with quality-controlled annotations and sample validation. These bimodal samples were obtained from 3554 audio clips and 2568 lyrics, also annotated and available for single-modality experiments. The dataset is publicly available\footnote{Available at: \url{https://zenodo.org/records/13939205}} and is annotated in two ways, based on Russell's circumplex plane \cite{russell1980CircumplexModelAffect}: i) four emotion quadrants; ii) and continuous arousal-valence (AV) values.

Even though this is not yet a large-scale dataset, to our knowledge, it is the largest publicly available and quality-controlled bimodal static MER dataset, as discussed in Section \ref{sec:dataset_review}. Moreover, besides emotion quadrants and AV values, the dataset provides annotations with multiple emotion tags and genre labels for each song, making it suitable for other problems besides static MER.

To create the dataset, we followed a semi-automatic protocol based on AllMusic\footnote{AllMusic is “a popular music database that provides professional reviews and metadata for albums, songs and artists” \cite{hu20082007MirexAudio}. URL: \url{https://www.allmusic.com/}} emotion tags and manual validation. This considerably accelerates a fully manual annotation stage while promoting annotation quality. 
 
To validate the effectiveness of the dataset, we performed a comprehensive set of experiments using state-of-the-art classical Feature Engineering (FE), Machine Learning (ML), and Deep Learning (DL) approaches. These experiments targeted both quadrant classification and AV regression problems, using audio-only, lyrics-only, and bimodal data. The attained results and analysis confirm the viability of the proposed datasets for benchmarking further static MER studies. The best-performing model (a bimodal neural network combining audio and lyrics) attained an F1-score of 79.21\%.\par

The main contributions of this article are the following:
\begin{itemize}
  \item a novel, public, and quality-controlled bimodal audio-lyrics MER dataset, annotated with emotion quadrants, AV values, multiple emotion tags, and genre labels;
  \item a semi-automatic approach to construct the dataset;
  \item an extensive set of experiments comprising audio-only, lyrics-only, and bimodal classification and regression, to establish a baseline for benchmarking.
\end{itemize}

The document is organized as follows. Section \ref{sec:background_related_work} reviews the relevant background and related work regarding available audio, lyrics, and bimodal static MER datasets. Section \ref{sec:proposed_datasets} presents the proposed semi-automatic creation protocol, generation of Train-Validation-Test (TVT) splits, and contents of the dataset. Section \ref{sec:baseline_methods} describes the methodologies followed for evaluating the proposed datasets and establishing a baseline for benchmarking. The results and insights obtained are discussed in Section \ref{sec:results}. Finally, Section \ref{sec:conclusion} presents the main conclusions and final thoughts of this study.

\section{Background and Related Work}\label{sec:background_related_work}
The employed emotion model is a key factor in the creation of MER datasets. Therefore, we start this section with a review of common emotion taxonomies. We then provide an overview of the data collection and annotation approaches. Finally, we critically review current static MER datasets, highlighting the limitations that our proposed new dataset aims to address.

\subsection{Emotion Taxonomies} \label{sec:emotion_taxonomies}
Psychology researchers have long discussed how emotions can be represented and classified. This study has led to the proposal of several emotion taxonomies over the last century, which can be grouped into two major paradigms: categorical (or discrete) models and dimensional models. The two paradigms have been the subject of active research in emotional psychology, and each has its strengths and weaknesses \cite{harmon2017importance}, as discussed in the following paragraphs.
\\
\subsubsection{Categorical Models}\hfill \break
\indent 
Discrete emotion theories propose the existence of fundamental emotions that are universally shared among cultures. \cite {harmon2017importance}.
In the categorical paradigm, emotions are represented as a set of discrete categories or emotional descriptors, e.g., Ekman's basic emotions (comprising anger, disgust, fear, happiness, sadness, surprise) \cite{Ekman01051992}, Plutchik's emotion wheel (which presents eight primary emotions in a cyclical format: acceptance, anger, anticipation, disgust, joy, fear, sadness, surprise) \cite{PLUTCHIK19803}, or Hevner's adjective circle (where 67 emotion labels are organized into 8 clusters) \cite{hevner1936ExperimentalStudiesElements}, among others.

Categorical models might be organized into at least four groups: models based on basic emotions, music-specific models, data-driven models, and models defined by music platforms.

\textbf{Models based on Basic Emotions.} Models in this group are based on the theory that humans have a discrete and limited set of basic emotions that are universal and innate \cite{Ekman01051992}. However, the exact number and nature of these basic emotions remain subjects of debate, and it has been argued that emotional responses often arise from the interplay of two or more core emotions \cite{Geetha2024multimodal}. Additionally, the emotions represented in some of these models may not be well-suited to music. For example, Ekman's model aimed to encode facial expressions, and some of the employed categories may not be adequate for the musical case (e.g., disgust). In contrast, others usually associated with music are absent (e.g., calm) \cite{hu2011improving}.

\textbf{Music-specific Models.} In contrast, music psychologists such as Kate Hevner have proposed categorical models to specifically represent emotions in music \cite{hevner1936ExperimentalStudiesElements}. The argument in Hevner's adjective circle (mentioned above) is that it contains descriptors that are musically more plausible. However, the origin of such descriptors is less substantiated. Additionally, in the design of the model, only classical music was employed. Several authors have proposed updates to Hevner’s adjectives circle, adding new terms and reorganizing the clusters, e.g., \cite{Schubert2003Hevner}. 

\textbf{Data-driven Models.} Besides the previous models (proposed by psychology researchers), MER researchers have also made contributions. A notable example is the Music Information Retrieval EXchange (MIREX) emotion taxonomy \cite{hu2007ExploringMoodMetadata}. The model was derived from the song metadata provided by AllMusic. At that time, a total of 179 emotion tags were available. The authors proposed a data-driven approach that resulted in a categorical emotion model comprising 29 emotion words grouped into five clusters. This model has also faced some criticism. Being purely data-driven, it lacks support from psychology studies \cite{laurier2011AutomaticClassificationMusical}. Moreover, there is semantic and acoustic overlap between some clusters \cite{laurier2011AutomaticClassificationMusical} (based on the analysis of the corresponding MIREX dataset, discussed in Section \ref{sec:dataset_review}).

\textbf{Models defined by Music Platforms.}
Some music platforms propose their own lists of emotion tags. One example is AllMusic, which currently comprises 305 emotion adjectives\footnote{https://www.allmusic.com/moods}. This extensive list encompasses a wide range of emotional responses triggered by music. However, the employed emotion adjectives might be ambiguous \cite{yang2011emotion}. Also, although these emotion labels were created (and assigned to the songs in the platform) by professional editors \cite{hu2007ExploringMoodMetadata}, it is unknown whether emotion psychology studies were employed to validate them.

In summary, categorical models offer a simple and intuitive approach to distinguishing between different emotions \cite{huAdvancingMusicEmotion2025}. A key parameter in this paradigm is the number of emotion words represented. A limited set of words may not adequately capture the full range of emotional responses elicited by music \cite{yang2011emotion}. However, using a more extensive emotion lexicon may not solve the problem, as the language used to describe emotions might be ambiguous and vary significantly between individuals \cite{Juslin01092004}. 
\\
\subsubsection{Dimensional Models}\hfill \break
\indent 
The dimensional perspective sustains that the fundamental components of emotions are a limited set of dimensions. Here, emotions are identified based on their positions on a hyperplane with typically two or three axes, e.g., valence or arousal \cite{russell1980CircumplexModelAffect}.

In this paradigm, Russell’s circumplex emotion model \cite{russell1980CircumplexModelAffect} has gained particular acceptance in the MER community. Supporters of this idea suggest that emotional states arise from the combination of two distinct neurophysiological systems: one for valence (pleasure-displeasure, i.e., the polarity of emotion in terms of positive and negative states, also known as pleasantness) and another for arousal or activity (aroused-not aroused, also known as activity, energy, or stimulation level). Russell even claimed that valence and arousal are the “core processes” of affect, constituting the raw material or primitive of emotional experience \cite{russell1980CircumplexModelAffect}.  

\begin{figure}[h]
    \centering
    \includegraphics[width=0.7\linewidth]{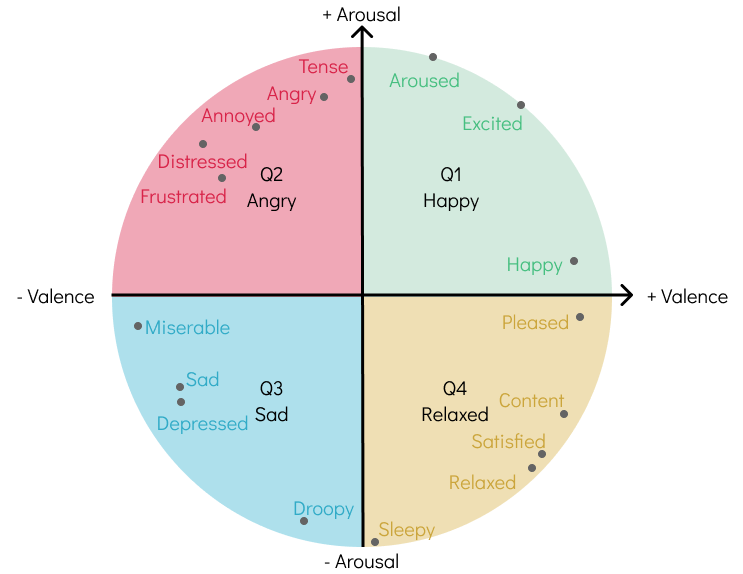}
    \caption{Russell's circumplex model of emotion as seen in \cite{louro2024ComparisonStudyDeep}.}
    \label{fig:russells_circumplex_model}
\end{figure}

The result, illustrated in Fig. \ref{fig:russells_circumplex_model}, is a two-dimensional plane, referred to as the arousal-valence (AV) plane. There, the X-axis represents valence and the Y-axis represents arousal, resulting in four quadrants that can be defined as: 1) positive valence and arousal, i.e., happy and energetic emotions such as excitement or enthusiasm (Quadrant 1 - Q1); 2) negative valence and positive arousal, i.e., frantic and energetic ones such as anxiety, fear or anger (Q2); 3. negative valence and arousal, i.e., melancholic and sad emotions such as depression (Q3); 4) and positive valence and negative arousal, representing calm and positive emotions such as contentment or serenity (Q4). 

Russell's circumplex model (and dimensional models in general) can be represented using a continuous or a discrete perspective, as follows. 

\textbf{Continuous Perspective.} In the continuous paradigm, there are no specific discrete emotion tags. Instead, emotions are regarded as a continuum. Thus, each point in the plane can represent a different emotion. For this reason, it is argued that the continuous paradigm better captures the complexity of the emotional space and entails lower ambiguity since no subjective tags are employed \cite{yang2008RegressionApproachMusic}. Nevertheless, in Russell's circumplex model, some important aspects of emotion might be obscured, leading to information loss. For example, emotions such as anger and fear (or boredom and melancholy) are closely placed in the AV plane but have distinct meanings. For this reason, a third dimension — dominance (or potency) — was proposed to distinguish such cases (e.g., fear and anger are close in the 2D AV plane but have opposite dominance, i.e., the former is submissive while the latter is dominant) \cite{yang2011emotion}. Another issue is that the continuous paradigm is less intuitive and increases the cognitive load on annotators, making it difficult to acquire consistent annotations. This inconsistency “heightens subjectivity, diminishes dataset reliability, and increases the potential for noise” \cite{huAdvancingMusicEmotion2025}. This problem is amplified when dominance is employed, which further increases model complexity and cognitive load; therefore, this dimension is not usually used. 

\textbf{Discrete Perspective.} In the discrete version of Russell's AV plane, different regions of the emotion plane represent different emotions, described by distinct emotion words (similarly to the previously described categorical emotion paradigm). Russell proposes several discrete emotions distributed across the AV plane (Fig. \ref{fig:russells_circumplex_model}). Besides this representation, Russell’s model may be simplified to comprise only four emotions, one for each quadrant, as previously mentioned. The relevance of this taxonomy was validated in a study by Laurier, where a semantic emotion space created from a community of users from the music social network Last.fm\footnote{\url{https://www.last.fm/}} was summarized into these four basic emotions \cite{laurier2011AutomaticClassificationMusical}. Although widely used in MER (mostly for its reduced operational complexity, e.g. \cite{panda2020NovelAudioFeatures}), the simplified 4-quadrant representation of Russell's AV plane has faced criticism, namely because it might neglect complex aspects of the emotional process and fail to capture the full range of emotional responses elicited by music \cite{huAdvancingMusicEmotion2025}. In summary, the discrete class of dimensional models entails the same advantages and drawbacks discussed above in the categorical emotion paradigm.

All in all, Russell's circumplex model provides a simple yet powerful representation of emotions. The two employed dimensions “provide a good balance between a parsimonious definition of emotion and the complexity of the study” \cite{yang2011emotion}. This is the main reason why it has received broad support from several music psychology studies \cite{JuslinSloboda2001} and has been adopted in several MER works, e.g., \cite{laurier2011AutomaticClassificationMusical, panda2020NovelAudioFeatures, wu2024popular}, despite the discussed limitations. 

\subsection{Data Collection and Annotation Approaches} \label{sec:data_collection_annotation}
After selecting an emotion model, data must be collected and annotated accordingly. This section starts with a brief overview of data collection tasks and then reviews data annotation strategies.

\hfill
\subsubsection{Data Collection, Pre-Processing and Cleaning} \hfill \break
\indent 
Researchers typically acquire audio clips from music platforms, e.g., AllMusic, Spotify, YouTube, and song lyrics based on lyrics web crawlers, e.g., lyrics.com, ChartLyrics\footnote{\url{http://www.chartlyrics.com/}}, MaxiLyrics\footnote{\url{http://www.lyricsmania.com/maxi_lyrics.html}} or MusixMatch\footnote{http://https://www.musixmatch.com/}. Such platforms usually offer Application Programming Interfaces (API).

In audio MER studies, a musical piece can be an entire song, a structural section of a song (e.g., chorus, verse, bridge, as in \cite{wu2024popular}), or a fixed-length clip (e.g., 30-second song excerpt, as in \cite{panda2020NovelAudioFeatures}).

For static MER, clips with approximately 30-second duration are commonly used to provide excerpts that convey a consistent predominant emotion throughout and to reduce the workload on human annotators \cite{hu20082007MirexAudio}. This duration is also used to comply with copyright issues. 

Such clips can be manually collected, which is a time-consuming process, albeit more controlled. Thus, they are often automatically extracted from the beginning or middle of the song. In other cases, song previews from music platforms like AllMusic are obtained using the provided API. However, the rationale for creating those excerpts is unclear, as those song previews either correspond to sections from the beginning, middle, and end of the song (or even sections containing only applause in live performances \cite{panda2020NovelAudioFeatures}. These processes typically require manual validation of the excerpts, as previous automatic strategies can result in clips that do not accurately represent the predominant emotion or may contain variations (or even noise, such as applause). For this reason, such clips should be validated by humans, although this is not always done, as seen in some datasets, e.g., \cite{huAdvancingMusicEmotion2025}.

The collected audio clips are then pre-processed, e.g., to standardize audio samples in terms of sampling rate, frequency, bit depth, and number of channels, and to remove noisy clips. 

Regarding song lyrics, it is essential to correct the text for orthographic errors and remove metadata and other descriptions inside the text (e.g., the name of the artist, title, and metadata about structural elements such as chorus, bridge, or verse) \cite{malheiro2018EmotionallyRelevantFeaturesClassification}.

\hfill
\subsubsection{Data Annotation} \hfill \break
\indent
After data collection, pre-processing, and cleaning, song annotation must be performed. To this end, different approaches have been employed in the literature, namely: 
\begin{itemize}
  \item Direct manual annotations, which comprise manual annotations via surveys \cite{yang2008RegressionApproachMusic} and manual annotations via annotation games \cite{kim2010MusicEmotionRecognition};
  \item Indirect annotations, e.g., where annotations are inferred from social tagging \cite{bertin-mahieux2011MillionSongDataset} and expert annotations provided by music platforms;
  \item Indirect annotations with human validation \cite{laurier2011AutomaticClassificationMusical, panda2020NovelAudioFeatures};
  \item Automatic annotations, e.g., via pre-trained models \cite{wu2024popular} or emotion lexicons \cite{cano2017MoodyLyricsSentimentAnnotated}.
\end{itemize}
This is the most challenging part, as emotion perception is inherently subjective and annotation is time-consuming \cite{yang2008RegressionApproachMusic}.

\textbf{Manual Annotations via Surveys.} 
Controlled manual annotations are typically performed via surveys. Each song is annotated by multiple subjects (typically more than 10) who are instructed to evaluate the emotional content of music in a controlled experimental setup. The process includes clearly defined emotion models, samples, and analysis of inter-annotator agreement \cite{aljanaki2017DevelopingBenchmarkEmotional}. Typically, the most prevalent opinion is selected. This process, coupled with an adequate quality-control protocol, has the potential to generate high-quality annotations. However, it requires hiring subjects, which can be expensive in terms of financial cost and time, besides being a tedious process, and, thus, error-prone \cite{kim2010MusicEmotionRecognition}. These difficulties are amplified for continuous AV annotations due to the previously mentioned cognitive load on evaluators, making it difficult to acquire consistent annotations and increasing the potential for noise \cite{huAdvancingMusicEmotion2025}. Moreover, for bimodal annotations, evaluators should make a mental effort to isolate the assessment of each source, e.g., ignoring the lyrics when annotating audio \cite{hu2007ExploringMoodMetadata, panda2020NovelAudioFeatures} and vice versa, as they may convey different emotions and affect the judgment of listeners. Several strategies can be implemented to reduce the impact of inaccurate annotations. These include discarding outlier evaluators and songs with low agreement among annotators using statistical tools such as average and standard deviation metrics \cite{malheiro2018EmotionallyRelevantFeaturesClassification} and inter-coder agreement metrics (e.g., Krippendorff’s alpha \cite{krippendorff2004ContentAnalysis}).

\textbf{Manual Annotations via Annotation Games.} 
To mitigate the effects of fatigue or lack of commitment associated with manual evaluation, an alternative method for annotating emotions in songs is through collaborative web games, also known as Games With A Purpose (GWAP), such as MoodSwings \cite{kim2010MusicEmotionRecognition}. The idea behind these games is to increase the commitment and motivation of annotators within a gaming context.

\textbf{Social Tagging.} 
To tackle the difficulties with direct manual annotations, social tags obtained directly from music social networks such as Last.fm can be employed \cite{laurier2011AutomaticClassificationMusical}. 
Compared to manual annotation, this method makes collecting ground truth data easier and faster. However, there are some problems with the obtained social tags: sparsity due to the cold-start problem and popularity bias, multiple spellings of tags, malicious tagging, or ad-hoc labeling techniques \cite{kim2010MusicEmotionRecognition}. For example, when a subject uses the tag “hate” on Last.fm, this might either mean that the song is about “hate” or that the person hates the song. As the study by Laurier concludes, “[manual] validation (...) is a necessary step to ensure the quality of the dataset”, because by “blindly follow[ing] the tags assigned by the community of Last.fm users, around 29\% of errors, on average, would have been introduced” \cite{laurier2011AutomaticClassificationMusical}.

\textbf{Expert Annotations from Music Platforms.} 
Compared to the previous approach, a potentially more robust alternative is to employ expert annotations provided by music platforms such as AllMusic \cite{hu2010ImprovingMusicMood}. For example, through the AllMusic web service, we can obtain song clips (excerpts with a duration of around 30 seconds, as mentioned above) and their corresponding emotion tags. However, as previously discussed, these tags are not part of any scientifically supported taxonomy. In addition, the annotation process in AllMusic is unclear: all we know is that the employed tags were “created and assigned to music works by professional editors” \cite{hu2007ExploringMoodMetadata}.
Additionally, the audio clips provided by the platform often contain noise (e.g., applause or silence) or segments that do not align with the assigned emotion labels. Hence, the provided music excerpts and the associated emotion tags require post-processing and validation \cite{panda2020NovelAudioFeatures}, as discussed above.

\textbf{Annotations from Music Platforms with Human Validation.} 
To partly overcome the described limitations, in \cite{panda2020NovelAudioFeatures}, we proposed a semi-automatic data collection and annotation strategy based on AllMusic annotations. The basic idea was to map the provided multi-label annotations of each song to a single emotion quadrant (according to Russell's model). To increase the robustness of the annotations, the quadrant proposed for each song was manually validated. In this article, we extend our original approach to address the bimodal case and to provide arousal-valence annotations, as described in Section \ref{sec:proposed_datasets}.

\textbf{Automatic Annotations.} 
Another possibility is to collect emotion annotations automatically. This strategy is applied, for example, in lyrics emotion datasets, where emotion lexicons such as ANEW \cite{bradley1999affective} or Warriner et al.'s \cite{warriner2013NormsValenceArousal}, which provide arousal-valence-dominance (AVD) ratings of English words \cite{cano2017MoodyLyricsSentimentAnnotated}. Nevertheless, the fact that human validation is not performed is a point of concern. Additionally, other approaches utilize pre-trained classification models to collect emotion annotations \cite{wu2024popular}. However, it is questionable whether labels provided by machine learning systems can be regarded as actual annotations.

\subsection{Current Static MER Datasets}\label{sec:dataset_review}
Over the years, several MER datasets have been introduced in the literature to address different MER problems, e.g., static MER \cite{panda2020NovelAudioFeatures}, multi-label MER \cite{huAdvancingMusicEmotion2025}, Music Emotion Variation Detection (MEVD) \cite{aljanaki2017DevelopingBenchmarkEmotional}, and music emotion induction \cite{zhang2018pmeo}, among others.

In the following, we provide an overview of static audio, lyrics, and bimodal audio-lyrics MER datasets as summarized in Table \ref{tab:static_MER_datasets}. In addition to static MER, we also review some multi-label MER datasets. Although these datasets do not directly address static MER, they could be adapted, for example, by mapping the employed emotion tags to Russell quadrants.
For each dataset, we provide a brief overview of the data collection and annotation process, highlighting its strengths and shortcomings.

\begin{table*}[!h]
	\caption{Static MER datasets (sorted by modality and year). \\4Q = Russell's 4 emotion quadrants; AV(D) = continuous AV(D) values; Bimodal = audio + lyrics; Game = manual annotations via annotation games; Manual = direct manual annotations; MP = expert annotations from Music Platforms; n.a. = not available / not applicable; ST = Social Tagging; VL = Variable Length.} \label{tab:static_MER_datasets}
    \centering
	\begin{tabular}{ccccccccc}
		\toprule
		\textbf{Dataset}	& \textbf{Year}	& \textbf{Modality}	& \textbf{Availability}	&\textbf{Size}	& 
        \begin{tabular}{c}
            \textbf{Sample}\\\textbf{Length}
        \end{tabular} & 
        \begin{tabular}{c}
        \textbf{Emotion}\\\textbf{Model}
        \end{tabular} & 
        \begin{tabular}{c}
        \textbf{Annotation}\\\textbf{Method}
        \end{tabular}\\ 
		\toprule
		\mbox{MIREX AMC \cite{hu20082007MirexAudio}}	& 2007	& Audio & Private & 600	& 30 sec	& \begin{tabular}{c} MIREX\\5 clusters \end{tabular} & Manual\\
            \midrule
		\mbox{Yang et al. \cite{yang2008RegressionApproachMusic}}	& 2008	& Audio	& Public & 195	& 25 sec	& AV  & Manual\\
            \midrule
      	\mbox{MagnaTagATune \cite{law2009EvaluationAlgorithmsUsing}}	& 2009	& Audio	& Public & 25,877	& 29 sec	& 27 labels  & Game\\
            \midrule
            \mbox{MSD \cite{bertin-mahieux2011MillionSongDataset}} & 2009	& Audio	& \begin{tabular}{c} Public (only\\audio features) \end{tabular} 
                & 1,000,000	& n.a.	& \begin{tabular}{c} Over 100\\labels
            \end{tabular}  & ST\\
      	\midrule
            \mbox{DEAM \cite{aljanaki2017DevelopingBenchmarkEmotional}}	& \begin{tabular}{c} 2013-\\2015 \end{tabular}	& Audio & Public 	& 1,744	& 45 sec	& AV  & Manual\\
            \midrule
            \mbox{\begin{tabular}{c} MSD \\Last.fm split \cite{choi2016AutomaticTaggingUsing} \end{tabular} }	& 2016	& Audio	& Private & 242,842	& n.a.	& \begin{tabular}{c} At least\\3 labels \end{tabular}  & ST	&\\ 
            \midrule
      	\mbox{4QAED \cite{panda2018MusicalTextureExpressivity}}	& 2018	& Audio	& Public & 900	& 30 sec	& 4Q  & 
        \begin{tabular}{c} MP with\\validation \end{tabular}	\\
            \midrule
      	\mbox{EMOPIA \cite{hung2021emopiamultimodalpoppiano}}	& 2018	& \begin{tabular}{c} Audio\\(+MIDI)
        \end{tabular}	& Public & 1,087	& VL	& 4Q  & Manual	\\
            \midrule
      	\mbox{MTG-Jamendo \cite{bogdanov2019mtg}}	& 2019	&    Audio	& Public & 55,609	& Full songs	& 59 labels  & ST\\    
            \midrule
      	\mbox{MuSe \cite{akiki2021muse}} 	& 2021	&    Audio	& \begin{tabular}{c}
                Public (only\\song IDs)
        \end{tabular} & 90,001	& 30 sec	& AVD  & Automatic\\            
        \toprule
      	\mbox{LED1 \cite{malheiro2018EmotionallyRelevantFeaturesClassification}}	& 2016	& Lyrics	& \begin{tabular}{c}
                    Public (links\\to lyrics)
                \end{tabular} & 180	& Full lyrics	& AV  & Manual\\
            \midrule
      	\mbox{LED2 \cite{malheiro2018EmotionallyRelevantFeaturesClassification}}	& 2016	& Lyrics	& \begin{tabular}{c}
                    Public (links\\to lyrics)
                \end{tabular} & 771	& Full lyrics	& 4Q  & Manual	\\
            \midrule
       	\mbox{MoodyLyrics 
\cite{cano2017MoodyLyricsSentimentAnnotated}}	& 2017	& Lyrics	& \begin{tabular}{c}
                    Public (links\\to lyrics)
                \end{tabular} & 2,595	& Full lyrics	& 4Q  & Automatic	\\ 
            \toprule
      	\mbox{Laurier et al. \cite{laurierMultimodalMusicMood2008}}	& 2008	& Bimodal	& Private & 1000	& 
            \begin{tabular}{c} 30 sec\\Full lyrics        
        \end{tabular} & 
         	4Q  & \begin{tabular}{c}
                ST with \\validation
        \end{tabular}	\\
        \midrule
        \mbox{Hu \& Downie \cite{hu2010ImprovingMusicMood}}	& 2010	& Bimodal	& Private & 5296	& \begin{tabular}{c} Full songs\\Full lyrics        
        \end{tabular}	& \begin{tabular}{c}
                18 emotion \\clusters
        \end{tabular}  & ST	\\
        \midrule
        \mbox{Delbouys et al. \cite{delbouys2018MusicMoodDetection}}	& 2018	& Bimodal	& Private & 18644	& \begin{tabular}{c} 30 sec\\Full lyrics        
        \end{tabular}	& AV  & ST	\\        
        \midrule
      	\mbox{Music4All \cite{santana2020music4all}}	& 2020	&    Bimodal	& \begin{tabular}{c}
                    \begin{tabular}{c} Upon\\request        
        \end{tabular}
                \end{tabular} & 109,269	& \begin{tabular}{c} 30 sec\\Full lyrics        
        \end{tabular} 	& AV  & Automatic\\    
        \midrule
      	\mbox{Popular Hooks \cite{wu2024popular}}	& 2024	&     \begin{tabular}{c} Bimodal \\ (+Video+MIDI) \end{tabular}	& Public & 38,694	&  \begin{tabular}{c} VL\\Full lyrics        
        \end{tabular}	& 4Q  & Automatic\\            
        \midrule
      	\mbox{Hu et al. \cite{huAdvancingMusicEmotion2025}}	& 2025	&    Bimodal	& \begin{tabular}{c} Public (only\\song IDs) \end{tabular} & 169,148	&  \begin{tabular}{c} Full songs\\Full lyrics        
        \end{tabular}	& 12 labels  & Automatic\\                    
        \bottomrule
	\end{tabular}
\end{table*}

\hfill 
\subsubsection{Audio-only Datasets} \hfill \break
\indent 
Initial attempts to propose benchmarks involved challenge-related databases, such as the MIREX Audio Mood Classification (AMC) dataset \cite{hu20082007MirexAudio} (for static MER). The MIREX dataset contains 600 audio clips (30 seconds each), equally distributed across the 5-cluster MIREX emotion taxonomy (previously described). The audio clips were carefully annotated by 15 human volunteers. However, several issues have been identified: i) the defined emotion taxonomy is not grounded in psychology studies; ii) some of the defined emotion clusters show semantic and acoustic overlap \cite{laurier2011AutomaticClassificationMusical}. Finally, the dataset is not publicly disclosed and can only be accessed within the MIREX challenge.

Another challenge-related contribution is the Database for Emotional Analysis in Music (DEAM) \cite{aljanaki2017DevelopingBenchmarkEmotional}, which resulted from the successive benchmarks for the 2013, 2014, and 2015 MediaEval Emotion in Music tasks. This dataset contains 58 full-length songs and
1,744 excerpts of 45 seconds (the ones considered for static MER) annotated with continuous arousal-valence (AV) values. Each audio clip was annotated by at least 10 subjects, and several quality-control procedures were implemented, including measuring annotator consistency and filtering out evaluators with low-quality annotations. Unlike the MIREX AMC dataset, DEAM is publicly available.

Apart from the above tentative benchmarks, Yang et al. presented one of the first public audio datasets \cite{yang2008RegressionApproachMusic}. The dataset comprised 25-second excerpts from 195 popular songs (representing the predominant emotion present, mainly the chorus) taken from Western, Chinese, and Japanese albums. The fully manual annotation process involved 253 subjects, who labeled ten random samples with AV values. The final AV values were obtained by averaging all annotations. The dataset quality was deemed acceptable based on the test-retest study, where the annotation process was repeated two months after the initial annotation. However, the dataset is very imbalanced. For example, only 12\% of all samples belong to the second quadrant.\par

The MagnaTagATune is a sizable music tagging dataset, aiming at multi-label classification and annotation through a gamification process \cite{law2009EvaluationAlgorithmsUsing}. A total of 25,877 samples from 5223 full songs are provided, accompanied by 29-second audio clips, with 188 unique tags (freely assigned by the players) across them, ranging from common high-level descriptors such as genre, emotion, and era to instruments and specific performing techniques (e.g., plucking). The list contains 27 emotion tags, e.g.,  happy, relaxed, joyful, angry, romantic, nostalgic, etc. According to \cite{choi2016AutomaticTaggingUsing,bogdanov2019mtg}, the dataset suffers from limitations, including noisy labels and an unbalanced tag distribution. Moreover, there is no standard data split for benchmarking, leading to inconsistency in the results reported using this dataset \cite{bogdanov2019mtg}.

The Million Song Dataset (MSD) is another example of a music tagging dataset. It was specifically designed to address the limited size of available datasets, and it remains the largest dataset in the MIR field. Data annotation is based on the collection of tags of multiple types, including emotion labels, provided by the users of the platforms such as Last.fm. A total of 522,366 tags are included. The exact number of emotion tags is unknown, but we estimate that it contains hundreds. A drawback of the MSD dataset is that it only provides audio features and metadata, so researchers typically collect song previews from external platforms. As different studies employ different song clips, distinct results are reported, making benchmarking difficult. Moreover, it suffers from the previously mentioned limitations of approaches based on social tags, namely the lack of tag validation and the associated ambiguities.

Choi et al. \cite{choi2016AutomaticTaggingUsing} developed another dataset focused on music tagging, based on a subset of the MSD, hereafter referred to as the MSD Last.fm split. It contains 242,842 samples obtained using track identifiers from the MSD. The dataset only includes samples with metadata that contains at least 50 unique general-purpose tags. As for emotion labels, the complete list is not disclosed, although three tags are mentioned: sad, happy, and chill. Despite being supposedly publicly available, the audio files cannot be acquired from the original provider, 7digital, as their API is no longer operational \cite{kim2023BiasedMSD}. 

Regarding our team's efforts, the most recent dataset released to the public is the 4-Quadrant Audio Emotion Dataset (4QAED) \cite{panda2020NovelAudioFeatures}. As previously mentioned, the creation of this dataset relied on collecting data and expert annotations from AllMusic. The main idea was to map the provided multi-label annotations of each song to an emotion quadrant (according to Russell's model) based on the ANEW English lexicon \cite{bradley1999affective}. The dataset comprises 900 audio samples with accompanying metadata, evenly distributed across the four quadrants. The sample data consists of 30-second song excerpts, their respective categorical labels (emotion quadrants and the original AllMusic tags), and 1714 extracted features. {As previously mentioned, the quadrant proposed for each song was manually validated.

MTG-Jamendo \cite {bogdanov2019mtg} is another large-scale music tagging dataset. It comprises 55,609 full audio tracks with 195 tags freely assigned by content uploaders of the Jamendo platform\footnote{https://jamendo.com/}. The list of tags comprises 95 genres and 59 emotions and themes, among others. Five random data splits (training, validation, and testing) are provided to permit benchmarking of the results. However, as in other large-scale datasets, the labels might be noisy and ambiguous since no manual validation was performed. Moreover, the dataset is significantly unbalanced.

The EMOPIA dataset \cite{hung2021emopiamultimodalpoppiano} is a particular dataset in this category, since it contains 1,087 variable-length audio clips (from 387 full songs), paired with MIDI samples. The clips were carefully annotated by four subjects. Moreover, song metadata and listener demographic information are provided. However, the dataset is restricted to pop piano music. 

The previously described large-scale datasets, such as MagnaTagATune, MSD, MSD Last.fm split, and MTG-Jamendo, all addressed multi-label emotion classification in the context of music tagging. These datasets employed emotion tags freely assigned by users. More recently, in 2021, the Musical Sentiment Dataset (MuSe) \cite{akiki2021muse} was introduced. This is a large-scale dataset specifically targeting static MER. The MuSe dataset includes arousal-valence-dominance (AVD) annotations for 90,001 songs, derived from user tags on Last.fm, incorporating 279 emotion labels from AllMusic as seeds. Similar to \cite{panda2020NovelAudioFeatures}, the emotion tags are mapped to the AVD cube using the previously mentioned English lexicon by Warriner et al. \cite{warriner2013NormsValenceArousal}, with the advantage that a weighted average of AVD values is calculated based on weights for each tag from Last.fm. However, it is important to note that audio clips are not directly available. Instead, the dataset provides Spotify IDs for 61,630 tracks, enabling the automatic retrieval of song previews, albeit with the limitations previously mentioned regarding the validity of the audio clips. Furthermore, according to Bogdanov et al. \cite{bogdanov2019musav}, only 41,021 30-second previews were accessible via the Spotify API (which was closed to public access by the end of 2024). Most importantly, no human manual validation has been conducted on the MuSe dataset. Therefore, the issues related to annotations based on social tagging remain unaddressed, which can lead to noisy annotations, as previously discussed. This is reflected in the low R2 scores obtained for arousal and valence (0.143 and 0.089, respectively), as discussed in \cite{bogdanov2019musav}, since the authors of the MuSe dataset did not provide a baseline.

\hfill
\subsubsection{Lyrics-only Datasets} \hfill\\
\indent 
Compared to audio-based MER approaches, fewer works have employed song lyrics, as reflected in the smaller pool of available datasets. Here, we describe lyrics-only datasets. Other datasets containing song lyrics were created in the context of bimodal MER, and will be described afterwards.

The largest dataset consisting solely of song lyrics is MoodyLyrics \cite{cano2017MoodyLyricsSentimentAnnotated}. It comprises 2595 lyrics annotated with Russell quadrants. It covers a high number of unique artists, and it is nearly balanced (although the first quadrant is over-represented). The construction of this dataset was fully automatic, by using emotion lexicons such as ANEW \cite{bradley1999affective} to map the aggregate AV values of all words in a song to each quadrant. However, as previously mentioned, the absence of human validation might be a point of concern. 

Our team developed two lyrics-only datasets (termed LED, for Lyrics Emotion Datasets), as presented in \cite{malheiro2018EmotionallyRelevantFeaturesClassification}, with a total of 942 lyrics. The first consisted of 180 manually annotated samples, and the second followed a semi-automatic annotation process akin to 4QAED, resulting in 771 samples. Some drawbacks of this dataset include its small size and slightly unbalanced quadrant distribution.


\hfill
\subsubsection{Bimodal Audio-Lyrics Datasets} \hfill\\
\indent 
The first bimodal dataset we are aware of is the quality-controlled 1000-song dataset created by Laurier et al. \cite{laurierMultimodalMusicMood2008}. It uses Last.fm as the source for audio and lyrics samples. The same platform provides the employed emotion tags, which are mapped to the four Russell quadrants and manually validated by 17 human subjects. 

Hu \cite{hu2010ImprovingMusicMood} created one of the first bimodal audio-lyrics datasets, which aimed at multi-label emotion classification. It includes 5296 audio and lyrics retrieved from the Last.fm platform, along with their emotion tags. Similar tags were clustered into larger groups, resulting in 18 emotion categories that formed a data-driven emotion taxonomy (not validated by music psychology studies). Manual validation is not mentioned, and its quality cannot be assessed since it is private.

Another contribution is the sizable 18644-song dataset by Delbouys et al. \cite{delbouys2018MusicMoodDetection}, which used the MSD as its source. The dataset was annotated with continuous AV values by automatically mapping Last.fm tags into the AV space using Warriner et al. 's lexicon of English words and their corresponding AVD values \cite{warriner2013NormsValenceArousal}. 30-second excerpts were extracted from the complete songs (no exact details about the procedure were provided). The annotations were heavily biased towards audio, leading to possible conflicts with the emotional content of the lyrics, and, as was the case for Hu's dataset, its quality cannot be assessed because it is private. Adding to the lack of manual validation, the low regression results suggest that the samples and annotations are noisy.

In 2020, the Music4All dataset was released \cite{santana2020music4all}. This large-scale dataset comprises 109,269 songs, annotated with AV values obtained automatically using the Spotify API (where the energy feature serves as a surrogate for arousal). The dataset is available upon request. Since the annotation relies on pre-trained models, this automatic process may introduce noise, as previously discussed and specifically confirmed for Spotify's valence and energy features \cite{panda2021does}. For the audio content, 30-second excerpts were automatically extracted from the middle of the song without any human validation. Additionally, lyrics were automatically obtained via MusixMatch, but there was no human validation to verify whether the emotions perceived in the acoustic and textual components matched, as discussed earlier. 

Popular Hooks, a fully accessible multimodal dataset, was introduced \cite{panda2021does}. The dataset comprises 38,694 hooks (i.e., memorable sections of songs) accompanied by synchronized audio, lyrics, music videos, and MIDI files. In addition to emotion annotations based on Russell's quadrants, a notable feature of this dataset is the inclusion of musical metadata such as song structure and tonality, as well as genre information. The emotion quadrant annotations were generated automatically using pre-trained models (one for each data source), and their accuracy was assessed through a user study involving 80 randomly selected samples from the dataset. Despite the authors' efforts to validate their automatic annotation approach, the small size of the validation sample might not be sufficient to ensure the quality of the annotations. Furthermore, no baseline was provided using the entire dataset. It is also important to note that the dataset is heavily unbalanced, with approximately 85\% of the data falling under Q1 and Q2.

Recently, Hu et al. \cite{wu2024popular} developed a large-scale bimodal dataset to address multi-label emotion classification. This dataset includes 169,148 complete songs from the NetEase Cloud Music platform, a freemium music streaming service based in China. A notable limitation of the dataset is that it only provides song IDs, which poses a challenge since the platform is exclusively available in China. In terms of emotion classification, the platform uses 12 predefined emotion tags: exciting, fresh, healing, happy, lonely, missing, nostalgic, quiet, romantic, relaxing, sentimental, and touching. The emotion labels for each song are gathered from user playlists. Although this process is automatic and lacks human validation, it appears reliable as it reflects the collective emotional consensus of multiple users. However, the employed taxonomy is limited in scope, as it does not cover a significant portion of the emotional spectrum; particularly, none of the tags fall within the second quadrant of Russell's circumplex model. The lyrics for the songs were obtained using the song IDs. Unlike other bimodal datasets, the authors took care to clean the data by removing irrelevant information, such as timestamps. However, similar to other datasets, there was no human verification to confirm if the emotions perceived in the acoustic and textual components aligned, as discussed previously. 

\hfill
\subsubsection{Discussion} \hfill\\
\indent 
Based on the analysis of the previous MER datasets, we highlight the following key points:
\begin{itemize}
    \item There is a significant diversity in the employed emotion taxonomies. However, Russell's circumplex model is the most common approach, whether it is represented through discrete quadrants or continuous AV values. All other models lack validation from music psychology studies.
    \item Most of the reviewed datasets exhibit a good variety of genres and styles.
    \item Several large-scale datasets have been proposed; however, all of them are annotated using social tagging or automatic methods without human validation, resulting in a lack of quality control.
    \item As expected, datasets that include quality control are generally smaller due to the need for thorough human validation. One of the largest datasets that incorporates such quality control is DEAM, which contains 1,744 audio clips.
    \item None of the bimodal datasets performs human validation of the annotations.
    \item Furthermore, none of the bimodal datasets verify whether the emotions perceived in the acoustic and textual components align with each other. It remains unclear whether participants annotated songs based on audio, lyrics, or both. As previously mentioned, audio and lyrics should be annotated separately to evaluate their specific contributions to music emotion recognition.
    \item While most of the reviewed datasets are publicly accessible, some of the large-scale datasets only provide song IDs. Even if it is possible to acquire the audio clips, validation is still necessary, as discussed earlier.
    \item Some reviewed datasets, particularly recent large-scale datasets, do not provide baselines for assessing feasibility and establishing a basis for future comparison.
\end{itemize}

In light of this analysis, we propose a new dataset in the following section to address some of the limitations identified in current static MER datasets.


\section{Proposed Dataset} \label{sec:proposed_datasets}
The proposed MERGE dataset comprises audio, lyrics, and bimodal modalities, enabling both single- and bimodal research. Each modality includes two variations: i) complete, i.e., all songs without any balancing; ii) balanced, focusing on even distribution across both quadrant and genre, following the protocol established by Panda et al.  \cite{panda2020NovelAudioFeatures}. The datasets, along with their metadata and extracted features (audio and lyrics), are publicly available\footnote{Available at: \url{https://zenodo.org/records/13939205}}. 

The remainder of this section describes the building process and contents of the dataset. Before that, we present a set of requirements to consider during the dataset creation process. \par

\subsection{Requirements for MER Datasets}
After reviewing the MER datasets, we have established the following requirements for the MERGE dataset:

\begin{itemize}
    \item[] \textbf{R1. Validated taxonomy}:  The dataset should be based on psychologically validated taxonomies. For simplicity, a reduced set of consensual emotional terms should be employed. From the discussion in Section \ref{sec:background_related_work}, Russell’s circumplex model appears as the most promising approach, whether through discrete quadrants or continuous AV values. Although the previously described study by Laurier \cite{laurier2011AutomaticClassificationMusical} concludes that Russell quadrants may adequately summarize emotion tags employed by Last.fm users, this simplified approach might fail to capture the complexity of the emotional space. Therefore, we propose using a taxonomy that combines both discrete Russell quadrants and continuous AV values.
    \item[] \textbf{R2. Variety and balance}: Datasets should be varied, balanced, and not limited to a single musical genre, style, or era.
    \item[] \textbf{R3. Careful annotation}: As discussed, datasets indirectly annotated from social tags or automatic systems should be validated by humans. In addition, noisy samples (audio and lyrics) should be discarded or cleaned.
    \item[] \textbf{R4. Reduced ambiguity}: At least good annotator agreement should be achieved, minimizing the mentioned ambiguity issues. This would lead to datasets with reasonably clear emotions, a key need at the current stage of MER research.
    \item[] \textbf{R5. Separate annotation between audio and lyrics}: When creating bimodal MER datasets (containing audio and lyrics), care should be taken to isolate the two sources in the annotation process so that the impact of each modality can be properly assessed.
    \item[] \textbf{R6. Public availability}: It is necessary that the datasets be public to permit a comparative analysis of different methods.
    \item[] \textbf{R7. Large size}: sizable datasets are required to exploit ML and DL solutions better.
    \item[] \textbf{R8. Provision of a baseline}: A baseline should be provided to assess the feasibility of the dataset and set the stage for future comparative analysis.
\end{itemize}

We also defined two additional secondary requirements:

\begin{itemize}
    \item[] \textbf{S1. Metadata for a wide range of research works}:  Besides emotion annotations, datasets should provide metadata such as genre, artist, album, year, and complete emotion tags. These would make the dataset relevant for the broader Music Information Retrieval (MIR) field and might be useful for later, more advanced tasks such as multi-label emotion classification.
    \item[] \textbf{S2. Semi-automatic construction process}: Probably, the main difficulty with the previous primary requirements is that at least part of the annotation process must involve manual human validation. This calls for semi-automatic construction approaches, reducing the resources needed to build a sizable dataset, as discussed below.  
\end{itemize}

\subsection{Creation Protocol}
We guided the creation of the new dataset by the above requirements.  Algorithm \ref{alg:dataset_creation} describes the dataset creation procedure (adapted and improved from our previous work \cite{panda2020NovelAudioFeatures}). We outline the key concepts of the proposed algorithm in the following sections.

After gathering audio clips from AllMusic using the provided API\footnote{Available audio samples and corresponding metadata were retrieved through \url{https://tivo.stoplight.io/docs/music-metadata-api}.}, a key step of our approach is the mapping of AllMusic emotion tags (curated by AllMusic experts) to Russell’s quadrants. To this end, we employ Warriner’s adjectives list \cite{warriner2013NormsValenceArousal}, which contains a list of 13915 emotion adjectives (in English) with affective ratings in three dimensions: arousal, valence, and dominance (AVD). We then map each song to a point in the AV plane by averaging the emotion tags based on Warriner's scores. To minimize ambiguity, songs located near the center of the plane, specifically in the interval [-0.2, 0.2] (on a [-1, 1] scale), are excluded. Additionally, we maximize genre variability in each quadrant.

Following the audio collection, their corresponding lyrics are retrieved from platforms such as lyrics.com, Chart-Lyrics, MaxiLyrics, and MusixMatch. In this process, lyrics could not be found for some of the audio samples. The lyrics were then cleaned through a series of operations. These include correcting spelling errors, eliminating lyrics that are not in English, removing lyrics with less than 100 characters, and removing metadata text,
among others. Additionally, the lyrics were complemented according to the corresponding audio. This means that repetitions of the chorus in the audio are added to the lyrics. Similarly, metadata defined in the lyrics (e.g., [Chorus x2]) implies adding one more instance of the chorus to the lyrics. After making these additions, the lyrics are then checked for any remaining cases of these patterns and eliminated. This process is described in greater detail in \cite{malheiro2018EmotionallyRelevantFeaturesClassification}.\par

A crucial step in our approach is the manual validation of the acquired songs, specifically in terms of the assigned quadrants and the quality of the audio clips and lyrics. To achieve this, we conducted a blind inspection of the candidate set. Participants received sets of randomly distributed audio clips and song lyrics and were asked to annotate them according to Russell’s quadrants. If any sample was of poor quality (such as being noisy, containing claps, or being silent), it was discarded. A song was retained if the annotated emotion quadrant aligned with the quadrant derived from mapping the AllMusic emotion tags; otherwise, it was discarded.  Similar to the method described in \cite{laurier2011AutomaticClassificationMusical}, we consider a song valid if at least one annotator confirms the tag.  This step is crucial because it reduces the cognitive burden associated with a fully manual annotation process. By ensuring that expert annotations are carefully obtained, we can validate these with minimal human resources. A total of 8 participants were involved in the validation process. 

The bimodal dataset is created from validated audio and lyrics datasets, including songs that correspond to the same audio and lyrics quadrants. These datasets are referred to as the “complete” datasets. To ensure balance, the “balanced” audio, lyrics, and bimodal datasets are formed by removing samples from the quadrants that are more heavily represented. This approach results in equally represented quadrants while also maintaining genre balance. Moreover, AV values are also provided (on a [-1, 1] scale.

In addition to the procedure described, the original 4QAED and LED datasets served as the foundation for creating the MERGE dataset. Whenever possible, lyrics were obtained for the audio-only samples from the 4QAED dataset, while audio was retrieved for the lyrics-only samples from the LED dataset.

The following paragraphs will discuss the resulting number and distribution of samples across quadrants for each dataset.

\begin{algorithm*}
\caption{Dataset creation algorithm.} \label{alg:dataset_creation}
\noindent\begin{minipage}{\textwidth} 
\begin{multicols}{2}
1. Gather songs and emotion data from AllMusic services.
    \begin{itemize}
        \item[] 1.1. Retrieve the list of 289\footnote{When we began creating this dataset, AllMusic offered 289 emotion labels; the current total is now 305, as previously mentioned.} emotion tags, $E$, using the AllMusic API.
        \item[] 1.2. For each emotion tag gathered, $E_i$, query the API for the top 10000 songs related to it, $S$.
    \end{itemize}
2. Bridge the tags from AllMusic with Warriner’s list.
    \begin{itemize}
        \item[] 2.1. For each emotion tag, $E_i$, retrieve the associated AV (arousal, valence) values from Warriner’s dictionary of English words. If the word is missing, remove it from the set of tags, $E$.
        \item[] 2.2. Using the retrieved AV values, map each emotion tag, $E_i$, onto one of the four Russel's quadrants.
        \item[] 2.3. Assign a quadrant to each song, $S_i$, based on the quadrant where the majority of the emotion tags, $E_i$, fall.
    \end{itemize}
3. Perform data pre-processing and filtering to reduce the massive amount of gathered data to a more balanced but still sizable set, $FS$.
    \begin{itemize}
        \item[] 3.1. Filter ambiguous songs (where a dominant emotional quadrant is not present). 
            \begin{itemize}
                \item[] 3.1.1. For all the songs in $S_i$, calculate the average arousal and valence values of all the emotion tags gathered, $E_i$.
                \item[] 3.1.2. If the average value of valence or arousal is in the range [-0.2, 0.2] (on a [-1, 1] scale), remove the song from the dataset.
            \end{itemize}
        \item[] 3.2. Remove duplicated or very similar versions of the same songs by the same artists (e.g., different albums) by using approximate string matching against the combination of artist and title metadata.
        \item[] 3.3. Eliminate songs without genre information. This ensures that the algorithms that maximize genre diversity can function correctly.
    \end{itemize}
4. Generate a subset, $GS$, maximizing genre variability in each quadrant.\\
5. Obtain the manually validated audio dataset, $ASV$.
    \begin{itemize}
        \item[] 5.1. Distribute all the songs in the set $GS$ equally among all team members.
        \item[] 5.2. For each song, $GS_i$, validation and annotation are performed according to Russell’s quadrants.
            \begin{itemize}
                \item[] 5.2.1. Verify that the song is valid (e.g., does not contain clapping, noise, or silence) and that the emotion present in the song is not ambiguous. 
            \end{itemize}
    \end{itemize}
6. Retrieve the lyrics dataset, $LS$, corresponding to the validated audio clips, $AVS$, from the following platforms: lyrics.com, ChartLyrics, MaxiLyrics, and MusixMatch, leading to the lyrics dataset (instrumental songs will be discarded from the lyrics dataset ).\par
7. Perform lyrics cleaning, e.g., spell checking, removal of non-English lyrics, removal of metadata, etc.\par
8. Obtain the manually validated lyrics dataset, $LSV$.
    \begin{itemize}
        \item[] 8.1. Distribute all the songs in the $LS$ set equally among all team members.
        \item[] 8.2. For each song, $LS_i$, perform validation and annotation of the song according to Russell’s quadrants.
            \begin{itemize}
                \item[] 8.2.1. Verify that the lyrics file is well structured, belongs to the correct audio clip, and that the emotion in the file is not ambiguous. 
                
            \end{itemize}
        \end{itemize}
9. Define the bimodal dataset, $Bm$, by keeping only the songs where audio and lyrics annotations match.
    \begin{itemize}
        \item[] 9.1. For each song, $ASV_i$ and $LSV_i$, if the annotated audio and lyrics quadrants match, the song is added to the bimodal dataset; otherwise, the song will be discarded (but present in the audio subset with a given quadrant and in the lyrics subset with a different quadrant).
    \end{itemize}
10. Create the final complete and balanced audio, lyrics, and bimodal datasets. 
    \begin{itemize}
        \item[] 10.1. The above $ASV$, $LSV$, and $Bm$ datasets form the complete sets, containing annotations for discrete quadrants and continuous AV values (on a [-1, 1] scale).
        \item[] 10.2 From the datasets in 10.1, obtain balanced datasets, $ASV_b$, $LSV_b$, and $Bm_b$, respectively, by discarding samples from the more represented quadrants, respecting genre balancing.
    \end{itemize}

\end{multicols}
\end{minipage}
\end{algorithm*}

\subsection{Dataset Description}
\label{sec:datasets_description}
The resulting datasets are hereafter termed \textit{MERGE Audio} (AC for the complete variation and AB for the balanced variation), \textit{MERGE Lyrics} (LC and LB), and \textit{MERGE Bimodal} (BC and BB) and are summarized in Table \ref{tab:datasets}.

\begin{table}[H] 
	\caption{Datasets used for evaluation with respective sample distribution.\label{tab:datasets}}
    \centering
	\begin{tabular}{cccccc}
		\toprule
		\textbf{Dataset}	& \textbf{Q1}	& \textbf{Q2}	& \textbf{Q3}	& \textbf{Q4}	& \textbf{Total} \\
		\midrule
		\mbox{MERGE Audio Complete}		& 875	& 915	& 808	& 956	& 3554\\
		\mbox{MERGE Audio Balanced}		& 808	& 808	& 808	& 808	& 3232\\
		\mbox{MERGE Lyrics Complete}	& 600	& 710	& 621	& 637	& 2568\\
		\mbox{MERGE Lyrics Balanced}	& 600	& 600	& 600	& 600	& 2400\\
  		\mbox{MERGE Bimodal Complete}   & 525	& 673	& 500	& 518	& 2216\\
		\mbox{MERGE Bimodal Balanced}	& 500	& 500	& 500	& 500	& 2000\\
		\bottomrule
	\end{tabular}
\end{table}

In short, MERGE AC contains 3554 samples, while MERGE AB includes 3232 (808 per quadrant). For lyrics, MERGE LC contains 2568 samples, while MERGE LB has 600 samples per quadrant (2400 samples in total). Finally, MERGE BC comprises 2216 samples, whereas MERGE BB contains exactly 2000 samples (500 samples per quadrant). As can be observed, the audio sets are larger since, as mentioned previously, retrieving lyrics from the corresponding songs was not always possible. Additionally, the bimodal dataset is smaller, as the annotated audio and lyrics quadrants do not always align.\par

In addition to audio clips and lyrics, each dataset includes individual metadata and train-validation-test (TVT) splits. The metadata file contains essential attributes such as the song identifier, title, artist, year, and genre tags, along with annotated quadrant and arousal-valence values (on a [-1, 1] scale). By providing this additional information, our datasets can be utilized for related tasks, such as music tagging (which employs audio and emotional labels) or era recognition. Regarding the provided AV values, it is important to bear in mind that these values were not manually validated; instead, they were obtained automatically from the mapping of the original emotion labels into the AV plane using Warriner's dictionary.

The TVT splits provided for each dataset come in two configurations (for training, validation, and testing): 70-15-15 and 40-30-30. These splits were created using a method designed to maximize quadrant balancing and genre distribution across each set. In addition to experiments that use k-fold cross-validation, we encourage researchers to use the proposed TVT splits rather than create their own. This approach ensures the reproducibility of results.

Finally, as we will show in the following sections, among the proposed datasets, experiments with MERGE Bimodal Complete (BC), using the 70-15-15 TVT split, typically yielded the highest F1-scores (which were comparable to cross-validation F1-scores). Also, this data split is well-suited for optimizing and quickly validating MER systems. For this reason, we propose the MERGE BC dataset and the 70-15-15 TVT split as the \textit{main profile} for future benchmarking.

\section{Baseline Methodologies and Evaluation Strategy} 
\label{sec:baseline_methods}
This section presents the overall evaluation strategy employed in this study, followed by descriptions of the baseline methodologies developed for this study. 

We performed various experiments to establish a baseline for benchmarking and provide a comprehensive evaluation of the proposed dataset. State-of-the-art approaches were employed for both quadrant classification and AV regression in each modality (audio-only, lyrics-only or bimodal), in essentially two categories: i) feature engineering and classical machine learning; ii) and deep learning. 

We note that the main purpose of these experiments is to assess the dataset's feasibility rather than to focus on the novelty of the employed models or to propose new methods. Our goal with the selected baseline methods is to establish a foundation for future, more advanced models. We are aware that other methods in the literature may achieve higher classification and regression scores.\par

\subsection{Evaluation Strategy}
As previously mentioned, we approach static MER as both classification and regression problems.

For these two problems, two evaluation methods were used to evaluate the performance of each methodology: stratified 10-fold cross-validation with 10 repetitions, and the two previously described TVT splits (70-15-15 and 40-30-30). Optimal hyperparameters were found through Bayesian search optimization. The obtained optimal parameters for all models are provided as supplementary materials.

Statistical significance tests were performed to compare the classification results from the proposed models and modalities on the relevant experiments. Differences are statistically significant for p $<$ 0.05. 

\subsection{Audio Classical ML}
\label{sec:audio_classical_ml_desc}
Our approach in \cite{panda2020NovelAudioFeatures} served as the basis for the conducted classical ML experiments. All songs are standardized, 
and features related to the eight standard musical dimensions (melody, harmony, rhythm, dynamics, expressivity, texture, and form) are extracted. Further details about features are available in \cite{panda2020NovelAudioFeatures}.\par

The ReliefF algorithm was then employed for feature ranking and selection. For classification, SVMs\footnote{Other classical ML approaches were evaluated (e.g., K-Nearest Neighbours and Random Forest), but SVMs achieved the best results.} were used, and their optimal hyperparameters were obtained through a Bayesian search. As for regression, Support Vector Regression (SVR) was employed. 
The kernel radial basis function (RBF) was selected, as earlier experimentation led to better results. This kernel requires tuning cost (C), set to [1e-6, 1500], and gamma, set to [1e-6, 100]. A logarithmic uniform step size was defined in both, allowing a higher number of smaller values to be tested.

Most of the implementations (SVMs and the Bayesian search algorithm used for model optimization) are provided by the scikit-learn Python library\footnote{\url{https://scikit-learn.org/}.}. ReliefF is an exception, being provided by the \textit{attrEval} function of the CORElearn R package\footnote{\url{https://www.rdocumentation.org/packages/CORElearn/}.}, for which no equivalent was found in Python.\par

In both evaluation methods, different numbers of features were tested to find the optimal feature set. This was performed independently for each dataset containing audio, as mentioned in Section \ref{sec:datasets_description}.

\subsection{Audio Deep Learning}
\label{sec:audio_dl_met}
Our audio DL baseline uses as inputs audio embeddings from the wav2vec2 model \cite{baevski2020Wav2vec20Framework}. These audio embeddings are processed to reduce them to a 1024-sized vector (depicted at the top of Figure \ref{fig:bimodal_dl_networks}), followed by two dense layers (at the right side of the same figure).



The wav2vec2 model family comprises pre-trained models specifically designed for speech recognition. The large version contains 960 hours of 16 kHz speech data, which we utilize in this study. The model processes raw audio through a multi-layer convolutional feature encoder and seven temporal convolutional blocks, outputting audio features. These features are then passed through a Transformer-like context network that employs dynamic convolutions for relative positional embeddings, enabling the extraction of information primarily related to acoustics and timbre.

The model produces a vector consisting of 1,024 values for each timestep across the 25 hidden layers of its context network. Since we focus on static emotion, we averaged the timestep information, resulting in a concise set of 1,024 values for each of the 25 layers (i.e., a 1024x25 matrix). This matrix is then fed to a convolutional 1D layer, followed by a flatten layer, to reduce the audio embedding to a 1024-sized vector.

The sample rate is lower compared to classical experiments to reduce the complexity of the model. It has also been stated that such reduction does not impact the model's performance \cite{pyrovolakis2022MultiModalSongMood}, as confirmed experimentally.\par

A Bayesian optimization strategy akin to the one used for the classical approach is applied using the Optuna library\footnote{\url{https://optuna.org/}.}. Hyperparameters tuned using this strategy include optimizer (Stochastic Gradient Descent (SGD), Adaptive Moment Estimation (Adam), and Adamax), learning rate, and batch size. Search spaces are defined as follows. The learning rate is searched between [1e-5, 1e-2], 
and batch size between [32, 256]. 

We implemented an early stopping strategy for the classifier methodologies to prevent the models from overfitting on the training data.
If the model's validation accuracy does not improve noticeably after 15 epochs, the optimization is stopped, and the next set of candidate hyperparameters is tested.

\subsection{Lyrics Classical ML}
The basis for the following machine learning experiments, which includes data pre-processing, feature selection, and the creation of classification and models, is described in \cite{malheiro2018EmotionallyRelevantFeaturesClassification}.\par

Regarding feature extraction, we use the features proposed in \cite{malheiro2018EmotionallyRelevantFeaturesClassification}, which are briefly divided into content-based (e.g., bags-of-words), stylistic (e.g., number of occurrences of nouns, adjectives, adverbs, slang words, etc.), song-structure (e.g., number of repetitions of the chorus and song title, etc.), and semantic features (e.g., features extracted from frameworks such as Synesketch\footnote{\url{https://github.com/parthenocissus/synesketch_v2.1/}.}, ConceptNet\footnote{\url{https://conceptnet.io/}.}, LIWC\footnote{\url{https://www.liwc.app/}.} and General Inquirer\footnote{\url{https://inquirer.sites.fas.harvard.edu/}.}, as well as features based on word dictionaries (gazetteers) related to each of Russell's emotion quadrants. 

As in audio, SVMs are used to create classification models, which are parameterized with an RBF kernel and tuned using Bayesian parameter search. 
We also employ the ReliefF algorithm for feature selection and ranking. \par

The optimization strategy presented in Section \ref{sec:audio_classical_ml_desc} (repeated 10-fold cross-validation and TVT) was also applied to the lyrics counterpart.  \par

\subsection{Lyrics Deep Learning}\par
Similar to audio, our lyrics DL baseline uses as inputs word embeddings from the Robustly Optimized BERT Pre-Training Approach (RoBERTa) pre-trained model \cite{liu2019RoBERTaRobustlyOptimized}. These word embeddings serve as inputs to a neural network with two dense layers (as depicted at the bottom of Figure \ref{fig:bimodal_dl_networks}).

RoBERTa employs self-attention and pre-training mechanisms on large text corpora to capture long-range emotional contexts. After experimentation, we found that encoding the full lyrics performed better than encoding individual verses. However, a caveat of obtaining RoBERTa's embeddings from the available HuggingFace implementation\footnote{Available at \url{https://huggingface.co/sentence-transformers/all-roberta-large-v1}.} limits the input to 512 characters, meaning that lyrics had to be truncated. 

As with audio, a Bayesian optimization strategy was also employed using the Optuna library, with the same search spaces. 

\subsection{Bimodal Classical ML}
We perform feature-level fusion to combine audio and lyrics in classical machine learning. The combined audio and lyrics features are fed to the ReliefF feature selection algorithm altogether, with the rest of the pipeline remaining unchanged.\par

One caveat that required special attention was the thousands of content-based features extracted from lyrics, which initially led to worse results when combined with the audio features. This was possibly due to the feature selection algorithm's inability to handle such high dimensionality. As such, in the developed bimodal classical machine learning approach, content-based lyrics features were discarded.\par


\subsection{Bimodal Deep Learning}\par
As shown in Figure \ref{fig:bimodal_dl_networks}, we employ an early fusion strategy with the previously discussed DL methodologies, akin to the ML approach described in the previous subsection. Audio embeddings are reduced as described in Section \ref{sec:audio_dl_met} and concatenated with the corresponding word embeddings for the same sample. The prediction is obtained using a simple two-layer dense network.\par

\begin{figure*}
    \centering
    \includegraphics[scale=0.20]{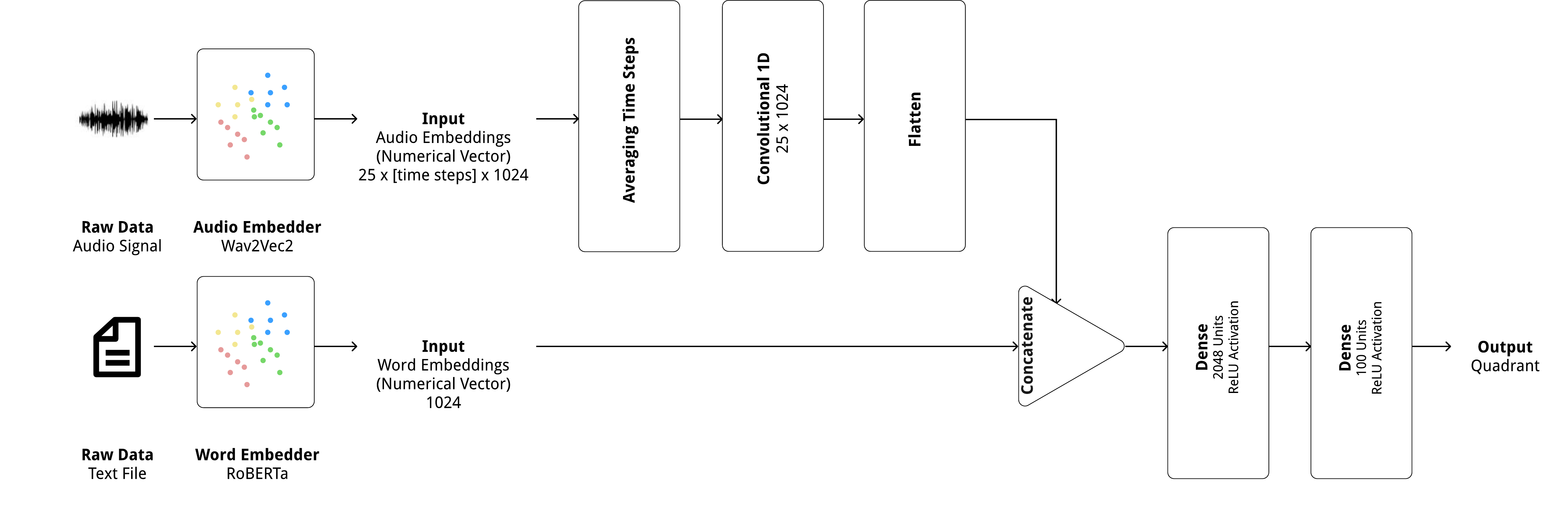}
    \caption{Architecture for bimodal DL experimentation. Audio and word embeddings are obtained for each sample. The former's output is further reduced before being concatenated with the latter. The concatenated vector is then classified with two dense layers.}
    \label{fig:bimodal_dl_networks}
\end{figure*}

The search spaces were mostly kept, except for the batch size, whose range was modified to [16, 128]. As observed in the lyrics DL experiments, the optimal batch size is relatively small compared to audio, so it made sense to adjust the corresponding range. \par

\section{Results and Discussion}\label{sec:results}
This section presents and discusses the results obtained for the audio, lyrics, and bimodal datasets after training models using the described baseline methodologies and evaluation strategies. For classification, F1-scores are presented; regarding regression, RMSE and R² scores are provided for arousal and valence, along with F1-scores obtained for classification based on regression (i.e., mapping the predicted AV values to the corresponding quadrant and measuring the resulting F1-score). As mentioned, we performed statistical significance tests in all comparisons, with a threshold set to p $<$ 0.05. Additionally, we utilize 4QAED and LED as baseline datasets for comparison purposes with the new datasets.

In the following subsections, several tables are presented to summarize the attained results. In those tables, CV stands for 10x10-fold Cross Validation, TVT for Train-Validate-Test (using 70-15-15 and 40-30-30 splits), CML for Classical ML, HF for Handcrafted Features,nAE for Audio Embeddings, WE for Word Embeddings, and DNN for Dense Neural Network. Regarding TVT, we only present F1-scores for compactness since we obtained similar recall and precision values. In those tables, several results are highlighted with bold font.

Besides the summary result tables mentioned, we also show the confusion matrices for the main profile, i.e., MERGE Bimodal Complete with TVT 70-15-15. Moreover, since the results obtained with the deep learning approach are typically higher than those obtained with the classical approach, we will present confusion matrices using the former.

\begin{table*}[!h]
    \caption{\textbf{Audio} Best Results \label{tab:all_dataset_metrics_audio}}
    \centering
        \begin{tabularx}{\textwidth}{@{} r C C *{5}{C} @{}}
        \hline
        \multirow{2}{*}{Dataset} & \multirow{2}{*}{Methodology} & \multirow{2}{*}{Approach} & Cross Val & & TVT 70-15-15 & & TVT 40-30-30\\ \cmidrule(lr{.2em}){4-4} \cmidrule(lr{.3em}){5-7} \cmidrule(l{.5em}){8-8}
        & & & F1-score & F1-score & R\(^2\) (A/V) & RMSE (A/V) & F1-score\\ 
        \hline
        \hline
        4QAED & \mbox{HF + SVM (CML)} & C & \textbf{71.71\%} $\pm$ 4.50  & - & - & - & -\\
        \hline
        \hline
        \multirow{3}{*}{MERGE} & \multirow{2}{*}{HF + SVM (CML)} & C & 71.01\% $\pm$ 2.31 & \textbf{70.12\%} & - & - & 66.38\% \\
        & & R & - & \textbf{70.18\%} & 0.479 / 0.364 & 0.281 / 0.375 & - \\
        Audio Complete & \multirow{2}{*}{AE + DNN (DL)} & C & 67.88\% $\pm$ 3.04 & 68.90\% & - & - & 64.69\% \\
        & & R & - & 66.71\% & 0.475 / 0.314 & 0.241 / 0.390 & - \\
        \hline
        \multirow{3}{*}{MERGE} & \multirow{2}{*}{HF + SVM (CML)} & C & 70.91\% $\pm$ 2.32 & 68.52\% & - & - & 68.20\% \\
        & & R & - & 67.47\% & 0.527 / 0.306 & 0.193 / 0.385 & - \\
        Audio Balanced & \multirow{2}{*}{AE + DNN (DL)} & C & 67.12\% $\pm$ 2.26 & 66.75\% & - & - & 67.79\% \\
        & & R & - & 68.03\% & \textbf{0.538} / 0.359 & 0.171 / 0.362 & - \\
        \hline
        \hline
        \multirow{3}{*}{MERGE} & \multirow{2}{*}{HF + SVM (CML)} & C & \textbf{71.03\%} $\pm$ 2.62 & 67.61\% & - & - & \textbf{69.22\%} \\
        & & R & - & 63.65\% & 0.498 / 0.284 & 0.192 / 0.403 & - \\
        Bimodal Complete & \multirow{2}{*}{AE + DNN (DL)} & C & \textbf{70.57\%} $\pm$ 3.18 & \textbf{70.84\%} & - & -  & 67.97\% \\
        & & R & - & \textbf{70.60\%} & 0.529 / \textbf{0.379} & \textbf{0.165} / 0.353 & - \\
        \hline
        \multirow{3}{*}{MERGE} & \multirow{2}{*}{HF + SVM (CML)} & C & 71.02\% $\pm$ 2.79 & 66.13\% & - & - & 67.70\% \\
        & & R & - & 68.76\% & 0.522 / 0.360 & 0.182 / \textbf{0.351} & - \\
        Bimodal Balanced & \multirow{2}{*}{AE + DNN (DL)} & C & 65.38\% $\pm$ 4.81 & 66.81\% & - & - & \textbf{68.84\%} \\
        & & R & - & 68.04\% & 0.531 / 0.255 & 0.168 / 0.404 & - \\
        \hline
        \end{tabularx}

        \begin{tablenotes}
        \footnotesize
        \item \textbf{Notes:} HF = Handcrafted Features, AE = Audio Embeddings.
        \end{tablenotes}
        
\end{table*}

\begin{table}[!t]
    \caption{\textbf{Audio} Confusion Matrix Results using the Main Profile (MERGE BC + TVT 70-15-15) with the DL approach) \label{tab:bimodalcomplete_conf_matrix_cv_audio_classical}}
    \centering
    \begin{tabular}{|c|c|c|c|c|c|}
        \cline{3-6}
        \multicolumn{2}{c|}{} & \multicolumn{4}{c|}{Predicted} \\
        \cline{3-6}
        \multicolumn{2}{c|}{} & Q1 & Q2 & Q3 & Q4 \\
        \hline
        \parbox[t]{2mm}{\multirow{4}{*}{\rotatebox[origin=c]{90}{Actual}}} & Q1 & \textbf{76.3\%} & 5.0\% & 5.0\% & 13.8\% \\
        \cline{2-6}
        & Q2 & 7.0\% & \textbf{93.0\%} & 0.0\% & 0.0\% \\
        \cline{2-6}
        & Q3 & 6.9\% & 0.0\% & \textbf{55.5\%} & 37.6\% \\
        \cline{2-6}
        & Q4 & 9.8\% & 2.0\% & 33.3\% & \textbf{54.9\%} \\
        \hline
    \end{tabular}
\end{table}

\subsection{MERGE Audio}
Table \ref{tab:all_dataset_metrics_audio} shows the overall results for the audio modality. Starting with a comparison between the baseline dataset (4QAED) and MERGE, similar results were obtained for the classical approach: 71.71\% for 4QAED\footnote{Although our implementation is identical to the original, we were unable to achieve the original 76.4\% score. This is a consequence of updates to the underlying feature extraction frameworks, leading to different values for some extracted features. For the sake of fair comparison between 4QAED and the novel datasets, we decided to report the results obtained under the same conditions. We will address this issue in future work.} and around 71\% for all MERGE datasets (in the cross-validation experiment). This suggests that the baseline and MERGE datasets have comparable complexity.  

Comparing the classical ML approach (relying on handcrafted audio features) and the DL approach, the former attains a maximum F1-score of 71.03\% (using the MERGE AC dataset with cross validation) and the latter tops at 70.84\% (using the MERGE BC dataset with the 70-15-15 TVT split). Although these top scores are comparable, we observe that the classical approach outperforms the DL approach in most cases. Results are only similar in the MERGE Bimodal Complete Dataset. This suggests that the employed audio embeddings have room for improvement compared to state-of-the-art acoustic features, such as through fine-tuning.


Regarding the influence of the size and imbalance of the new datasets, these factors showed little impact, as the results obtained for the four datasets (MERGE AC and AB, MERGE BC and BB) are similar. This is observed across all modalities covered in this study (audio, lyrics, and bimodal), as shown in the following subsections.

As for the standard deviation of the F1-scores for 10x10-fold CV, we can observe that they are low
(from 1.97\% to 4.81\%), which denotes low sensitivity to the defined folds. This is also observed in the lyrics and bimodal methodologies.

When compared with cross-validation (CV), TVT attains, in general, slightly lower but comparable results (for example, a top result of 71.03\% in CV against 70.84\% in TVT 70-15-15). This indicates the robustness of the proposed TVT splits and their feasibility for benchmarking, resulting in more straightforward and faster model training compared to cross-validation (CV). Comparing the two proposed splits, 70-15-15 outperforms 40-30-30 (a top F1-score of 70.84\% in the former against 68.84\% in the latter). This might result from the more extensive training set in the 70-15-15 split. 

Comparing the F1-scores using direct classification approaches with the ones from regression-based classification, we observe that the results attained by the regression-based strategy are around 2-3\% lower. This is to be expected considering the semi-automatic approach employed to obtain AV values for samples. However, this is not always the case, as results are slightly higher for the DL approach on the MERGE BC set, while the drop is only 1\% in the BB set.

Regarding R\(^2\) and RMSE values, as expected, valence performs significantly worse than arousal. Moreover, it fluctuates more across datasets, yielding R\(^2\) intervals between 0.255 and 0.379, compared to 0.475 and 0.538 for arousal. RMSE are also consistently higher for valence in comparison to arousal. Given the fact that the reference AV values were obtained automatically, we believe the obtained scores are acceptable.

Finally, the confusion matrix for the main profile is presented in Table \ref{tab:bimodalcomplete_conf_matrix_cv_audio_classical}. As can be observed, the model can accurately predict Q2, followed by Q1. However, despite our efforts to reduce ambiguity in the datasets, some confusion remains between Q3 and Q4, which leads to a lower score in these quadrants. This aligns with other studies in the literature that show the difficulty in distinguishing valence in low-arousal quadrants using only acoustical information \cite{panda2020NovelAudioFeatures}.

\begin{table*}[!h]
    \caption{\textbf{Lyrics} Best Results \label{tab:all_dataset_metrics_lyrics}}
    \centering
        \begin{tabularx}{\textwidth}{@{} r C C *{5}{C} @{}}
        \hline
        \multirow{2}{*}{Dataset} & \multirow{2}{*}{Methodology} & \multirow{2}{*}{Approach} & Cross Val & & TVT 70-15-15 & & TVT 40-30-30\\ \cmidrule(lr{.2em}){4-4} \cmidrule(lr{.3em}){5-7} \cmidrule(l{.5em}){8-8}
        & & & F1-score & F1-score & R\(^2\) (A/V) & RMSE (A/V) & F1-score\\ 
        \hline
        \hline
        LED & \mbox{HF + SVM (CML)} & C & \textbf{72.94\%} $\pm$ 4.42 & - & - & - & - \\
        \hline
        \hline
        \multirow{3}{*}{MERGE} & \multirow{2}{*}{HF + SVM (CML)} & C & 67.46\% $\pm$ 2.87 & 70.98\% & - & - & 64.95\% \\
        & & R & - & 59.18\% & 0.333 / 0.401 & 0.263 / 0.370 & - \\
        Lyrics Complete & \multirow{2}{*}{WE + DNN (DL)} & C & 68.82\% $\pm$ 5.17 & 71.40\% & - & - & 74.45\% \\
        & & R & - & 67.42\% & 0.366 / \textbf{0.569} & 0.205 / 0.298 & - \\
        \hline
        \multirow{3}{*}{MERGE} & \multirow{2}{*}{HF + SVM (CML)} & C & 67.48\% $\pm$ 2.49 & 69.25\% & - & - & 65.99\% \\
        & & R & - & 67.47\% & 0.358 / 0.387 & \textbf{0.193} / 0.385 & - \\
        Lyrics Balanced & \multirow{2}{*}{WE + DNN (DL)} & C & 68.01\% $\pm$ 3.79 & 74.15\% & - & - & 74.92\% \\
        & & R & - & 65.33\% & 0.311 / 0.492 & 0.213 / 0.332 & - \\
        \hline
        \hline
        \multirow{3}{*}{MERGE} & \multirow{2}{*}{HF + SVM (CML)} & C & \textbf{69.31\%} $\pm$ 3.27 & \textbf{71.31\%} & - & - & \textbf{69.57\%} \\
        & & R & - & 59.37\% & 0.350 / 0.389 & 0.255 / 0.391 & - \\
        Bimodal Complete & \multirow{2}{*}{WE + DNN (DL)} & C & \textbf{70.90\%} $\pm$ 2.86 & \textbf{75.05\%} & - & - & \textbf{74.99\%} \\
        & & R & - & 69.42\% & \textbf{0.365} / 0.556 & 0.191 / 0.305 & - \\
        \hline
        \multirow{3}{*}{MERGE} & \multirow{2}{*}{HF + SVM (CML)} & C & 66.96\% $\pm$ 3.35 & 69.50\% & - & - & 66.74\% \\
        & & R & - & 57.60\% & 0.351 / 0.377 & 0.260 / 0.369 & - \\
        Bimodal Balanced & \multirow{2}{*}{WE + DNN (DL)} & C & 69.34\% $\pm$ 4.59 & 72.83\% & - & - & 72.84\% \\
        & & R & - & 70.69\% & 0.350 / 0.552 & 0.196 / \textbf{0.299} & - \\
        \hline
        \end{tabularx}

        \begin{tablenotes}
        \footnotesize
        \item \textbf{Notes:} HF = Handcrafted Features, WE = Word Embeddings.
        \end{tablenotes}
\end{table*}

\begin{table}[!t]
    \caption{\textbf{Lyrics} Confusion Matrix Results using the Main Profile (MERGE BC + TVT 70-15-15) with the DL approach) \label{tab:bimodalbalanced_conf_matrix_cv_lyrics_dl}}
    \centering
    \begin{tabular}{|c|c|c|c|c|c|}
        \cline{3-6}
        \multicolumn{2}{c|}{} & \multicolumn{4}{c|}{Predicted} \\
        \cline{3-6}
        \multicolumn{2}{c|}{} & Q1 & Q2 & Q3 & Q4 \\
        \hline
        \parbox[t]{2mm}{\multirow{4}{*}{\rotatebox[origin=c]{90}{Actual}}} & Q1 & \textbf{76.5\%} & 2.9\% & 5.9\% & 14.7\% \\
        \cline{2-6}
        & Q2 & 2.9\% & \textbf{87.3\%} & 5.9\% & 3.9\% \\
        \cline{2-6}
        & Q3 & 2.8\% & 6.9\% & \textbf{76.4\%} & 13.9\% \\
        \cline{2-6}
        & Q4 & 25.6\% & 2.2\% & 13.3\% & \textbf{58.9\%} \\
        \hline
    \end{tabular}
\end{table}

\subsection{MERGE Lyrics}
Table \ref{tab:all_dataset_metrics_lyrics} shows the overall results for the lyrics modality. Starting with a comparison between the baseline dataset (LED) and MERGE, contrary to audio, the results in the new datasets using the classical approach slightly underperform the ones attained in the baseline dataset (from 72.94\% in LED\footnote{It is worth noting that the 73.6\% F1-score reported in \cite{malheiro2018EmotionallyRelevantFeaturesClassification} was obtained on the LED2 771-lyrics subset; here, we performed 10x10-fold CV on the combined LED1+LED2 datasets (180 + 771 = 942 lyrics), hence, the slight differences.} to a maximum of 69.31\% on MERGE BC). Despite the increased size, this suggests that the complexity of the novel lyrics datasets increased compared to LED.

Comparing the classical ML approach (relying on handcrafted text features) and the DL approach, the former attains a maximum F1-score of 71.31\% and the latter tops at 75.05\% (both using the MERGE BC dataset with the 70-15-15 TVT split). We observe that the DL approach outperforms the classical one in most cases. This suggests that the employed word embeddings can capture the emotional content of the lyrics more accurately than the handcrafted features. This is unsurprising, as these embeddings were trained on large amounts of text data. Additionally, fine-tuning the word embeddings may further improve the obtained F1-scores.

Additionally, we observe that higher scores were obtained in the lyrics-only experiment compared to the audio-only experiment (a maximum of 75.05\% for lyrics, against 70.84\% for audio, using the main profile). This may suggest a higher maturity level of the word embeddings in comparison to audio embeddings, which was expected.


As before, the new datasets' size and imbalance had little impact. Once again, the results attained for the four datasets (lyrics complete and balanced, bimodal complete and balanced) are similar.

Regarding the standard deviation of the F1-scores for 10x10-fold CV, we can again observe a reasonably low sensitivity to the data folds (from 2.49\% to 4.59\%). 

When compared with CV, TVT consistently attains higher scores (for example, a top result of 75.05\% in TVT against 71.31\% in CV), indicating its robustness. Once again, the 70-15-15 split outperforms the 40-30-30 split, although in a less notorious way (a top F1-score of 75.05\% in the former against 74.99\% in the latter).

Continuing the trend observed in the audio modality, regression-based classification methodologies underperformed compared to direct classification approaches. The difference is most noticeable in the classical approaches, which range from 10\% to 13\% lower F1-score. Embeddings-based methodologies appear more robust, showing at most a 9\% lower F1-score and 2\% at best.

In the R\(^2\) metric, as expected for lyrics, valence outperformed arousal prediction. Interestingly, the classical approach yields very similar results for arousal and valence in terms of R\(^2\). However, the fact that the RMS error was also higher seems contradictory. This may indicate that the present model does not accurately predict samples outside the mean of the data. As for variability, the results do not vary as much as in audio for either axis.

Finally, the confusion matrix for the main profile is presented in Table \ref{tab:bimodalbalanced_conf_matrix_cv_lyrics_dl}. As can be observed, the model can accurately predict Q2, followed by Q1 and Q4. Compared to audio, the F1-scores obtained for Q1 are comparable, while the ones for Q2 are lower (though still high, i.e., 93.0\% against 87.3\%). Regarding Q3, we observe a significant improvement (from 55.5\% in audio to 76.4\% in lyrics), although some confusion between Q3 and Q4 still remains. As for Q4, results are only slightly higher compared to audio (58.9\% against 54.9\%). However, while in audio the main confusion came from Q4 and Q3, in lyrics the main observed confusion stemmed from Q4 and Q1). This was expected, given the theoretical knowledge that lyrics are better at capturing valence, while audio is superior for arousal \cite{malheiro2018EmotionallyRelevantFeaturesClassification}.

\begin{table*}[!h]
    \caption{\textbf{Bimodal} Best Results \label{tab:all_dataset_metrics_bimodal}}
    \centering
        \begin{tabularx}{\textwidth}{@{} r C C *{5}{C} @{}}
        \hline
        \multirow{2}{*}{Dataset} & \multirow{2}{*}{Methodology} & \multirow{2}{*}{Approach} & Cross Val & & TVT 70-15-15 & & TVT 40-30-30\\ \cmidrule(lr{.2em}){4-4} \cmidrule(lr{.3em}){5-7} \cmidrule(l{.5em}){8-8}
        & & & F1-score & F1-score & R\(^2\) (A/V) & RMSE (A/V) & F1-score\\ 
        \hline
        \hline
        \multirow{3}{*}{MERGE} & \multirow{2}{*}{HF + SVM (CML)} & C & \mbox{\textbf{78.58\%} $\pm$ 2.47} & \textbf{77.98\%} & - & - & \textbf{75.90\%} \\
        & & R & - & 62.04\% & 0.320 / 0.433 & 0.256 / 0.377 & - \\
        Bimodal Complete & \multirow{2}{*}{AE + WE + DNN (DL)} & C & \mbox{\textbf{81.74\%} $\pm$ 2.69} & \textbf{80.53\%} & - & - & \textbf{78.69\%} \\
        & & R & - & 74.58\% & 0.505 / \textbf{0.528} & \textbf{0.169} / \textbf{0.305} & - \\
        \hline
        \multirow{3}{*}{MERGE} & \multirow{2}{*}{HF + SVM (CML)} & C & \mbox{77.34\% $\pm$ 2.41} & 77.28\% & - & - & 75.18\% \\
        & & R & - & 71.15\% & \textbf{0.545} / 0.412 & 0.189 / 0.320 & - \\
        Bimodal Balanced & \multirow{2}{*}{AE + WE + DNN (DL)} & C & \mbox{79.42\% $\pm$ 2.72} & 75.75\% & - & - & 77.97\% \\
        & & R & & 75.64\% & 0.484 / 0.508 & 0.176 / 0.310 & - \\
        \hline
        \end{tabularx}
        
        \begin{tablenotes}
        \footnotesize
        \item \textbf{Notes:} HF = Handcrafted Features, AE = Audio Embeddings, WE = Word Embeddings.
        \end{tablenotes}
\end{table*}

\begin{table}[!t]
    \caption{\textbf{Bimodal} Confusion Matrix Results using the Main Profile (MERGE BC + TVT 70-15-15) with the DL approach) \label{tab:bimodalcomplete_conf_matrix_cv_bimodal_classic}}
    \centering
    \begin{tabular}{|c|c|c|c|c|c|}
        \cline{3-6}
        \multicolumn{2}{c|}{} & \multicolumn{4}{c|}{Predicted} \\
        \cline{3-6}
        \multicolumn{2}{c|}{} & Q1 & Q2 & Q3 & Q4 \\
        \hline
        \parbox[t]{2mm}{\multirow{4}{*}{\rotatebox[origin=c]{90}{Actual}}} & Q1 & \textbf{79.3\%} & 3.7\% & 3.7\% & 13.4\% \\
        \cline{2-6}
        & Q2 & 3.1\% & \textbf{95.9\%} & 0.0\% & 1.0\% \\
        \cline{2-6}
        & Q3 & 1.4\% & 1.4\% & \textbf{80.0\%} & 17.2\% \\
        \cline{2-6}
        & Q4 & 13.3\% & 1.2\% & 21.7\% & \textbf{63.9\%} \\
        \hline
    \end{tabular}
\end{table}

\subsection{MERGE Bimodal}
Regarding the experiments using the bimodal datasets, Table \ref{tab:all_dataset_metrics_bimodal} summarizes the results achieved. 


When comparing the bimodal, audio-only, and lyrics-only approaches, results in the novel datasets for the classical approach show that the bimodal strategy significantly outperforms the best methods from the isolated modalities, as expected: the bimodal model attained a maximum F1-score of 78.58\%, against 71.03\% for audio and 69.31\% for lyrics, all on MERGE BC. The same happens for DL approaches, where the bimodal methodology reached 81.74\% (the top overall result achieved in all the experiments conducted in this study), against 75.05\% for lyrics-only and 70.84\% for audio-only.  



When comparing TVT splits, as before, the 70-15-15 split outperforms the 40-30-30 split. An F1-score of 80.53\% was achieved with 70-15-15  against 78.69\% in 40-30-30. When compared with CV, TVT attains again comparable results.

The obtained results for regression-based classification are overall lower, as expected. Again, deep learning approaches appear more robust than the classical counterparts, with the largest difference from the direct classification approach being 6\% F1-score on MERGE BC set when compared to the 16\% difference of the classical methodology on the MERGE BC. 

The best results for R\(^2\) and RMSE are observed in this modality. The highest attained R\(^2\) score was 0.545 for arousal (on the MERGE BB dataset) and 0.528 for valence (on MERGE BC), respectively.  The smallest error observed for both axes in terms of RMSE is also observed on MERGE BC, achieving 0.169 for arousal and 0.305 for valence. 

Finally, Table \ref{tab:bimodalcomplete_conf_matrix_cv_bimodal_classic} contains the confusion matrix for the main profile. As can be observed, the scores increased for all quadrants compared to the audio-only and lyrics-only solutions. Compared to audio, a particularly noticeable improvement was obtained for Q3 (from 55.5\% to 80.0\%). For Q4, the increase was not so dramatic, although still significant (from 54.9\% to 63.9\%). Compared to lyrics, the improvements in these quadrants were not so notable, but were still very relevant (3.6\% and 5\% improvements in Q3 and Q4, respectively). This reinforces the conclusion that most of the improvement in the classification of the lower arousal quadrants is due to the lyrics.

The previous results confirm the potential of bimodal approaches to reduce the confusion between low-arousal quadrants. Yet, the attained results show that there is plenty of room for improvement and that the separation between Q4 and Q3, and Q4 and Q1, is far from being solved \cite{panda2020NovelAudioFeatures}. 


\section{Conclusion} \label{sec:conclusion}
This article proposed a new bimodal audio-lyrics dataset. For each modality, both a complete and balanced variation are available. Two TVT splits were created and released alongside these datasets to enable fast experimentation and guarantee uniformity for all research works that employ them.\par

To validate the proposed datasets and data splits, we performed experiments using classical approaches (based on handcrafted features and standard ML algorithms) and DL methodologies (based on audio and word embeddings).

Based on the obtained results, we conclude that the proposed datasets (along with the related semi-automatic creation protocol) and TVT data splits are viable for MER benchmarking. In addition, the methods employed provide a solid baseline for comparison with future works using the MERGE dataset.

This responds to a critical need in this research area, particularly the bimodal dataset, which is the main contribution of this study. To the best of our knowledge, this is the largest publicly available and quality-controlled MER bimodal dataset. In this respect, the approaches employing the bimodal dataset outperformed audio-only and lyrics-only strategies, further confirming the importance of leveraging audio and lyrics information to resolve ambiguity. 

Moreover, among the proposed datasets, MERGE Bimodal complete typically led to the highest F1-scores. Also, the proposed data splits, especially the 70-15-15 strategy, are well-suited for optimizing and quickly validating MER systems. For this reason, we propose the MERGE BC dataset and the 70-15-15 TVT split as the \textit{main profile} for future benchmarking.

Additionally, the proposed datasets are designed for various research purposes. In addition to emotion quadrant and arousal-valence annotations, the datasets also include metadata like genre, artist, album, year, and complete emotion tags. These features could benefit a wide range of MIR research and advanced MER tasks, including multi-label emotion classification.

Due to the current dataset sizes, the DL-based methods used in this work have not yet fully utilized the potential of deep learning. Although current DL methodologies in the literature open up many exciting research paths, the need for extensive data is still an issue that needs to be addressed. In this respect, a preliminary study \cite{louro2024ComparisonStudyDeep} shows the promise of hybrid approaches. The combination of handcrafted features with deep neural networks outperformed traditional feature engineering and machine learning methods. Therefore, despite its (still) limited size, the MERGE dataset is a step toward unlocking the potential of deep learning solutions for MER.

The methods employed in this work aimed to establish a baseline for benchmarking future work. As such, there is plenty of room for improvement, e.g., exploiting the potential of hybrid feature engineering and deep learning approaches, advancing research on new emotionally relevant features (particularly for musical expressivity, texture, and form \cite{panda2020NovelAudioFeatures}), or novel deep learning architectures.

\section*{Acknowledgments}
This work is funded by FCT - Foundation for Science and Technology, I.P., within the scope of the projects: MERGE - DOI: 10.54499/PTDC/CCI-COM/3171/2021 financed with national funds (PIDDAC) via the Portuguese State Budget; and project CISUC - UID/CEC/00326/2020 with funds from the European Social Fund, through the Regional Operational Program Centro 2020. Renato Panda was supported by Ci2 - FCT UIDP/05567/2020.
 
\bibliography{bibliography}

\vskip -2\baselineskip plus -1fil

\begin{IEEEbiography} 
[{\includegraphics[width=1in,height=1.25in,clip,keepaspectratio]{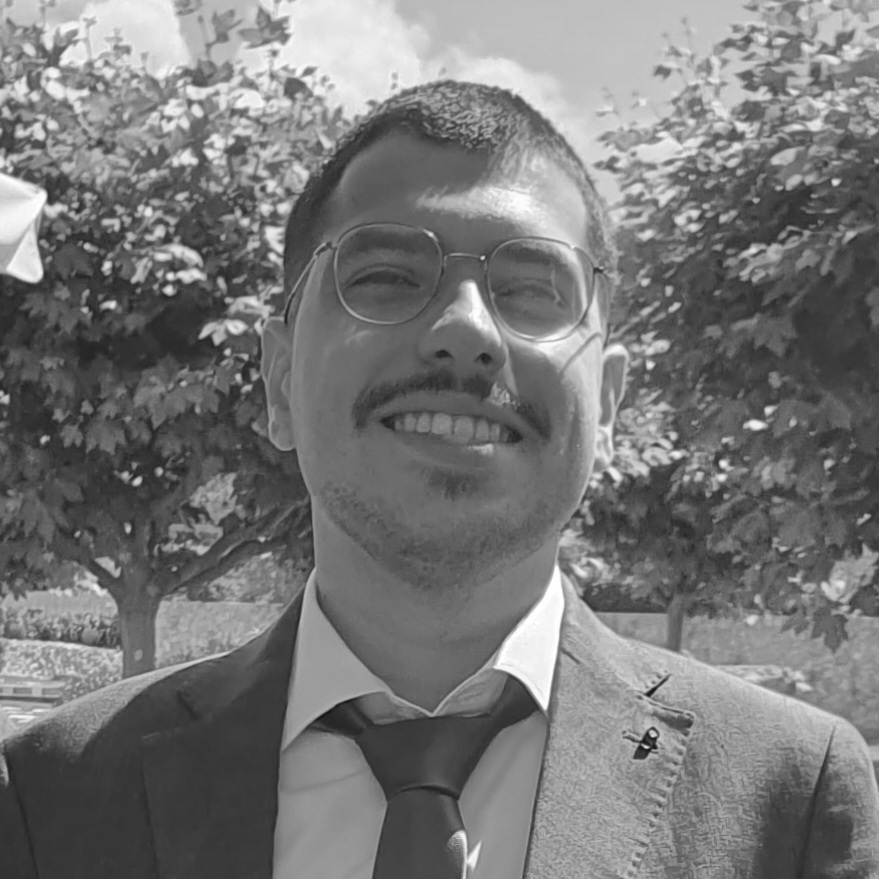}}]{Pedro Lima Louro} is a PhD Research Student at the University of Coimbra, where he also concluded is Masters degree, specializing in Intelligent Systems. He is a member of the Cognitive and Media Systems (CMS) research group at the Centre for Informatics and Systems of the University of Coimbra (CISUC). His main research interests include Music Information Retrieval (MIR), Music Emotion Recognition (MER), and Deep Learning.
\end{IEEEbiography}

\vskip -2\baselineskip plus -1fil

\begin{IEEEbiography}[{\includegraphics[width=1in,height=1.25in,clip,keepaspectratio]{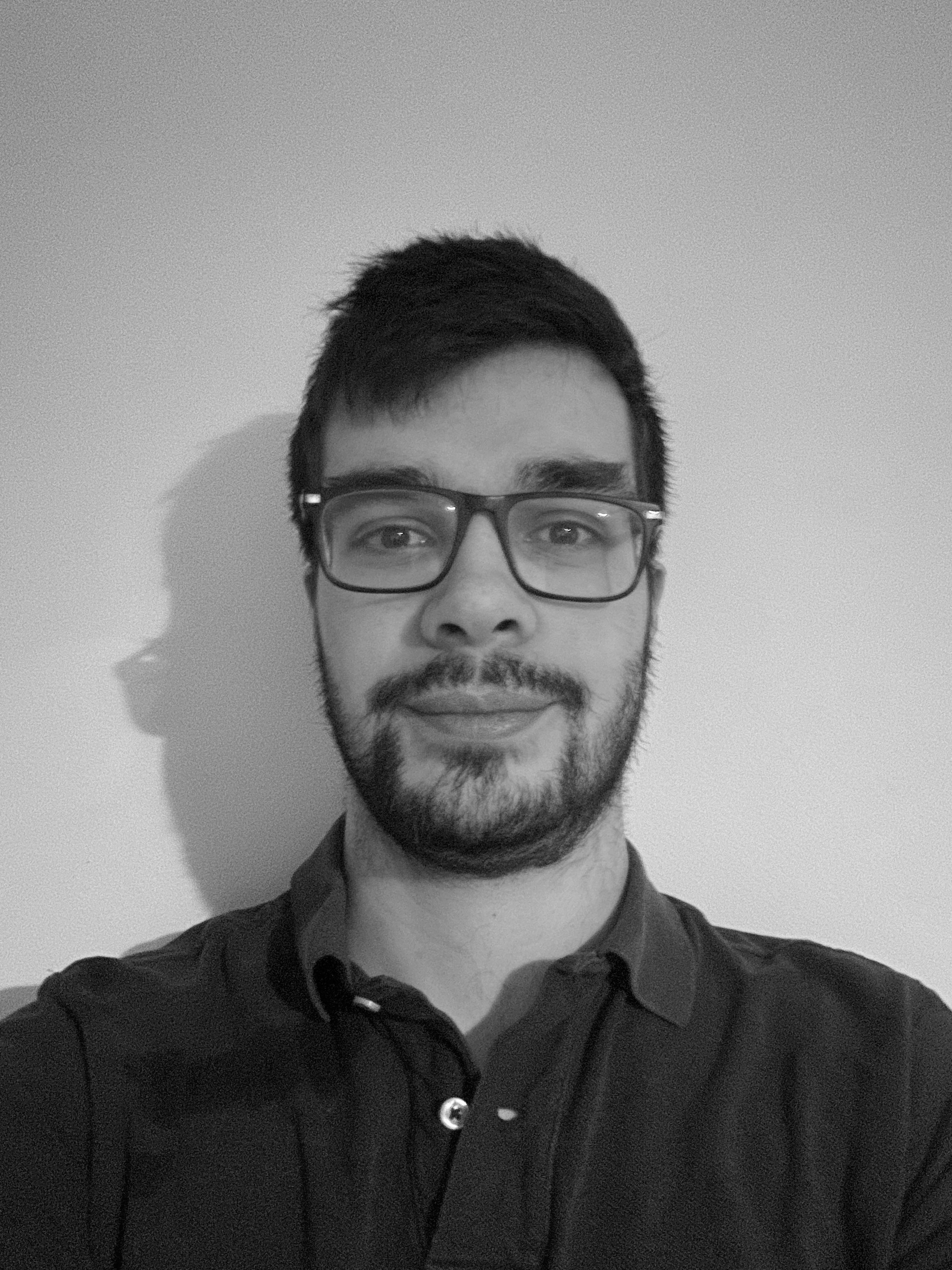}}]{Hugo Redinho} is an MSc from the University of Coimbra, specializing in Intelligent Systems, where he also concluded his Bachelor's degree in Informatics Engineering. He is a member of the Music Information Retrieval (MIR) research team at the Center for Informatics and Systems of the University of Coimbra (CISUC). His main research interests are related to Music Emotion Recognition (MER) and MIR.
\end{IEEEbiography}

\vskip -2\baselineskip plus -1fil

\begin{IEEEbiography}
[{\includegraphics[width=1in,height=1.25in,clip,keepaspectratio]{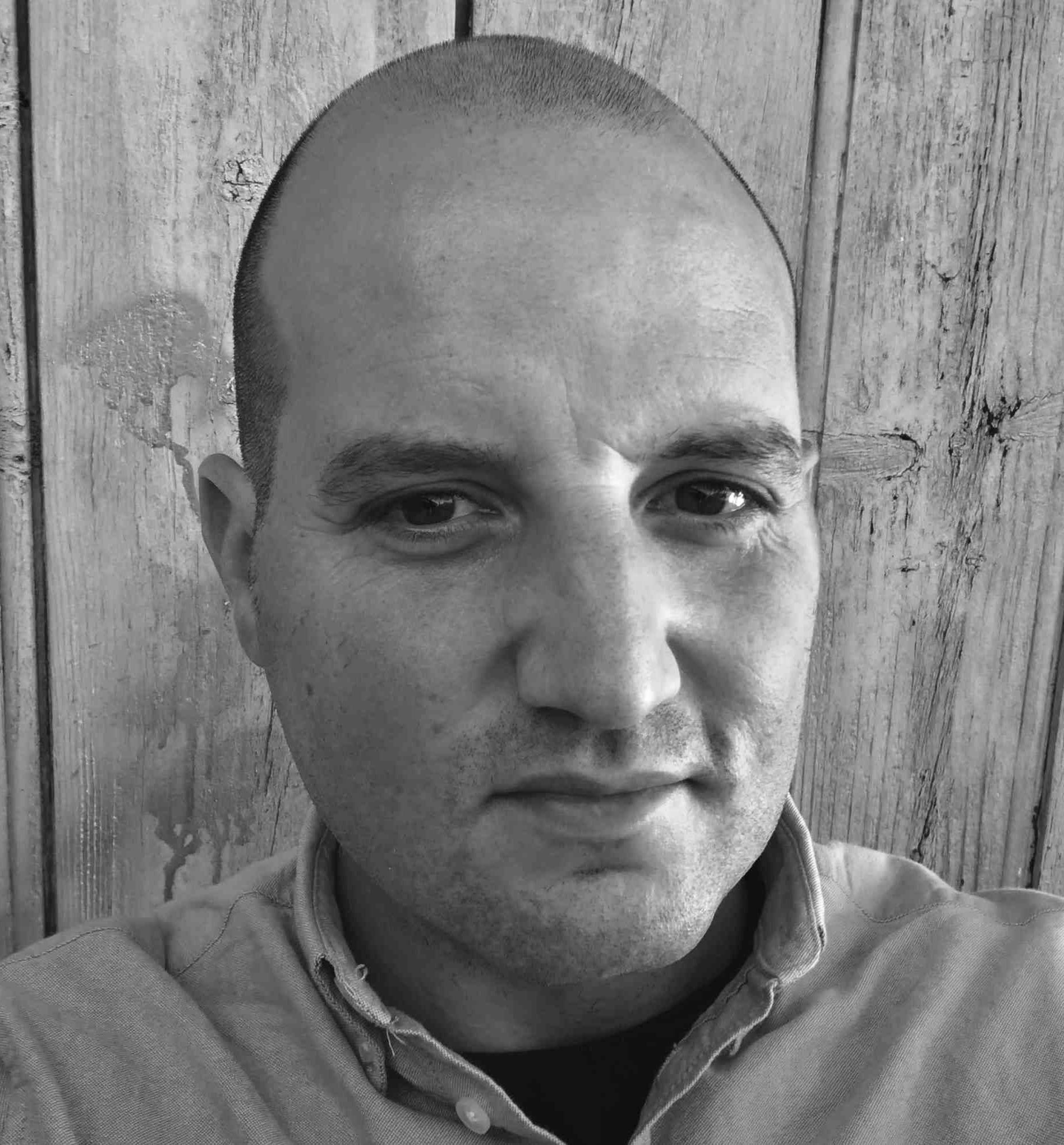}}]{Tiago Filipe Rodrigues Ribeiro} is a PhD Research Student at the Centre for Informatics and Systems of the University of Coimbra (CISUC). His research focuses on applying machine learning to analyze and generate emotions in music, with a particular emphasis on lyrics-based methods. He holds a Master’s degree in Data Science and a Bachelor’s degree in Electrical and Computer Engineering from the Polytechnic of Leiria, Portugal. He is particularly interested in the intersection of technology and humanities. 
\end{IEEEbiography}

\vskip -2\baselineskip plus -1fil

\begin{IEEEbiography}
[{\includegraphics[width=1in,height=1.25in,clip,keepaspectratio]{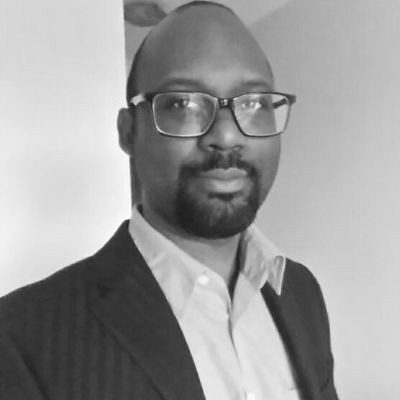}}]{Ricardo Santos} is a PhD Research Student at the Centre for Informatics and Systems of the University of Coimbra (CISUC). His research interests include Music Emotion Recognition, Deep Learning, and Large Language Models. Santos received his Master's in Computer Engineering from IPT/USP, Brazil.
\end{IEEEbiography}

\vskip -2\baselineskip plus -1fil

\begin{IEEEbiography}[{\includegraphics[width=1in,height=1.25in,clip,keepaspectratio]{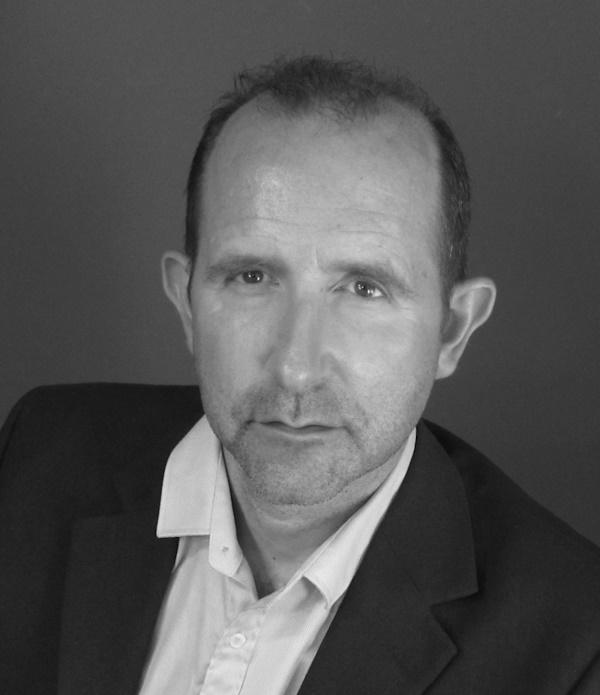}}]{Ricardo Malheiro} is a PhD from the University of Coimbra, where he also concluded his Master and Bachelor (Licenciatura - 5 years) degrees, respectively in Informatics Engineering and Mathematics. He is a Professor at the Polytechnic Institute of Leiria - School of Technology and Management. He is also a member of the Cognitive and Media Systems (CMS) research group at CISUC. His main research interests are Natural Language Processing, Detection of Emotions in Music Lyrics and Text, and Text/Data Mining.
\end{IEEEbiography}

\vskip -2\baselineskip plus -1fil

\begin{IEEEbiography}[{\includegraphics[width=1in,height=1.25in,clip,keepaspectratio]{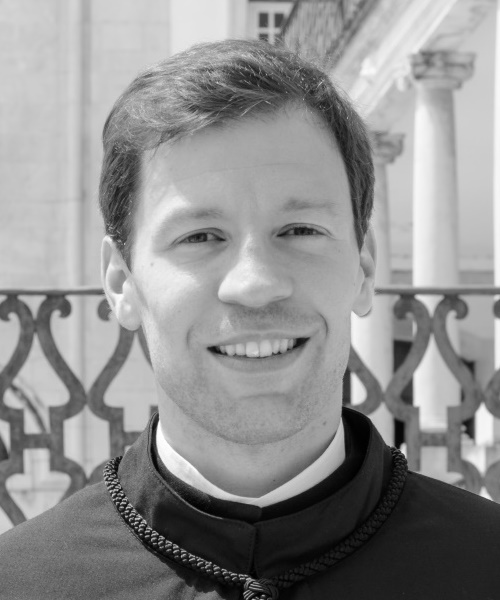}}]{Renato Panda} is an Assistant Researcher at Ci2, Polytechnic Institute of Tomar, Portugal. His main research interests are Music Emotion Recognition (MER) and Music Information Retrieval (MIR), as well as Applied Machine Learning and Software Engineering. He earned his PhD in Informatics Engineering from the University of Coimbra in 2019. Since then, he has been a member of the Cognitive and Media Systems group at the Centre for Informatics and Systems of the University of Coimbra (CISUC), where he remains actively involved.
\end{IEEEbiography}

\vskip -2\baselineskip plus -1fil

\begin{IEEEbiography}[{\includegraphics[width=1in,height=1.25in,clip,keepaspectratio]{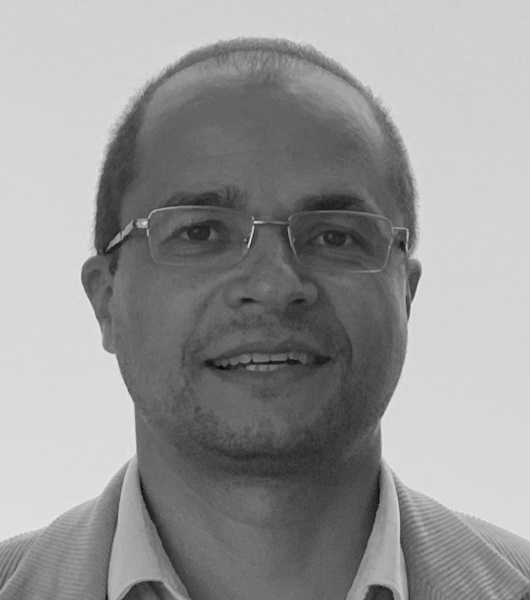}}]{Rui Pedro Paiva} is a Professor at the Department of Informatics Engineering of the University of Coimbra, where he concluded his Doctoral, Master and Bachelor degrees in 2007, 1999 and 1996, respectively. He is also a member of the CMS group at CISUC. His main research interests are in the areas of MIR and Health Informatics. The common research hat is the study of feature engineering, machine learning, and signal processing to analyze musical and bio signals.
\end{IEEEbiography}

\vfill

\end{document}